\documentclass[11pt,journal, onecolumn, draftcls]{IEEEtran}
\usepackage{amsmath,amssymb,latexsym}
\usepackage{graphicx}
\usepackage{subfigure}
\usepackage{epsfig}

\usepackage{psfrag}
\title{Joint Bandwidth and Power Allocation
with Admission Control in Wireless Multi-User Networks With and Without Relaying
}
\author{Xiaowen Gong, Sergiy A.~Vorobyov, and Chintha
Tellambura

\thanks{This work is supported in parts by the Natural Science and
Engineering Research Council (NSERC) of Canada and the
Alberta Ingenuity Foundation, Alberta, Canada.

The authors are with the Department of Electrical and Computer
Engineering, University of Alberta, 9107-116 St., Edmonton,
Alberta, T6G~2V4 Canada. Emails: {\tt \{xgong2, vorobyov, chintha
\}@ece.ualberta.ca}

{\bf Corresponding author:} Sergiy A.~Vorobyov, Dept. Elect. and
Comp. Eng., University of Alberta, 9107-116 St., Edmonton,
Alberta, T6G 2V4, Canada; Phone: +1 780 492 9702, Fax: +1 780 492
1811. Email: {\tt vorobyov@ece.ualberta.ca}. }
}

\begin{document}

\maketitle

\begin{abstract}
Equal allocation of bandwidth and/or power may not be efficient
for wireless multi-user networks with limited bandwidth and power
resources. Joint bandwidth and power allocation strategies for
wireless multi-user networks with and without relaying are
proposed in this paper for (i)~the maximization of the sum
capacity of all users; (ii)~the maximization of the worst user
capacity; and (iii)~the minimization of the total power
consumption of all users. It is shown that the proposed allocation
problems are convex and, therefore, can be solved efficiently.
Moreover, the admission control based joint bandwidth and power
allocation is considered. A suboptimal greedy search algorithm is
developed to solve the admission control problem efficiently. The
conditions under which the greedy search is optimal are derived
and shown to be mild. The performance improvements offered by the
proposed joint bandwidth and power allocation are demonstrated by
simulations. The advantages of the suboptimal greedy search
algorithm for admission control are also shown.
\end{abstract}


%\newpage

\section{Introduction}
One of the critical issues in wireless multi-user networks is
efficient allocation of available radio resources in order to
improve the network performance \cite{Julian}, \cite{Xiao}.
Therefore, power allocation strategies in wireless multi-user
networks have been extensively studied
\cite{Yates}--\cite{Chiang}. In practical wireless networks,
however, both the available transmission power of individual nodes
and the total available bandwidth of the network are limited and,
therefore, joint bandwidth and power allocation must be considered
\cite{Kumaran}--\cite{Maric}. Such joint bandwidth and power
allocation is important for both systems with and without
relaying. In the case of severe channel conditions in direct
links, relays can be deployed to forward the data from a source to
a destination in order to improve communication efficiency
\cite{Laneman}--\cite{Hong}.

There are numerous works conducted on the resource allocation in
wireless relay networks (see, for example,
\cite{Luo}--\cite{Xie}). Power allocation with the
decode-and-forward (DF) protocol has been studied in \cite{Luo}
under the assumption that transmitters only know mean channel
gains. In \cite{Havary}, a power allocation scheme that aims at
maximizing the smallest of two transceiver signal-to-noise ratios
(SNRs) has been studied for two-way relay networks. To optimize
effective capacity in relay networks, time/bandwidth allocation
strategies with constant power have been developed based on time
division multiple access/frequency division multiple access
(TDMA/FDMA) in \cite{Xie}. However, \cite{Luo}--\cite{Xie} as well
as most of the works on the resource allocation in wireless relay
networks consider the case of a single user, i.e., a single
source-destination pair.

Resource allocation for wireless multi-user relay networks has
been investigated only in few works
\cite{Senthuran}--\cite{Phan2}. Power allocation aiming at
optimizing the sum capacity of multiple users for four different
relay transmission strategies has been studied in \cite{Serbetli},
while an AF based strategy in which multiple sources share
multiple relays using power control has been developed in
\cite{Phan1}, \cite{Phan2}.

It is worth noting that the works mentioned above (except
\cite{Xie}) have assumed equal and fixed bandwidth allocation for
the one-hop links from a source to a destination. In fact, it is
inefficient to allocate the bandwidth equally when the total
available bandwidth is limited. Moreover, the problem of joint
allocation of bandwidth and power has never been considered for
wireless multi-user relay networks.

Various performance metrics for resource allocation in multi-user
networks have been considered. System throughput maximization and
the worst user throughput maximization are studied using convex
optimization in \cite{Julian}. Sum capacity maximization is taken
as an objective for power allocation in \cite{Serbetli}, while
max-min SNR, power minimization, and throughput maximization are
used as power allocation criteria in \cite{Phan1}.

In some applications, certain minimum transmission rates must be
guaranteed for the users in order to satisfy their
quality-of-service (QoS) requirements. For instance, in real-time
voice and video applications, a minimum rate should be guaranteed
for each user to satisfy the delay constraints of the services.
However, when the rate requirements can not be supported for all
users, admission control is adopted to decide which users to be
admitted into the network. The admission control in wireless
networks typically aims at maximizing the number of admitted users
and has been recently considered in several works. A single-stage
reformulation approach for a two-stage joint resource allocation
and admission control problem is proposed in \cite{Matskani},
\cite{Sidiropoulos3}, while another approach is based on user
removals \cite{Le}, \cite{Andersin}.

In this paper, the problem of joint bandwidth and power allocation
for wireless multi-user networks with\footnote{An earlier
exposition of this part of the work has been presented in
\cite{XiaoICASSP}.} and without relaying is considered, which is
especially efficient for the networks with both limited bandwidth
and limited power. The joint bandwidth and power allocation are
proposed to (i)~maximize the sum capacity of all users;
(ii)~maximize the capacity of the worst user; (iii)~minimize the
total power consumption of all users. The corresponding joint
bandwidth and power allocation problems can be formulated as
optimization problems that are shown to be convex. Therefore,
these problems can be solved efficiently by using convex
optimization techniques. The joint bandwidth and power allocation
together with admission control is further considered, and a
greedy search algorithm is developed in order to reduce the
computational complexity of solving the admission control problem.
The optimality conditions of the greedy search are derived and
analyzed.

The rest of this paper is organized as follows. System models of
multi-user networks without relaying and with DF relaying are
given in Section~\ref{sec_md}. In Section~\ref{sec_al}, joint
bandwidth and power allocation problems for the three
aforementioned objectives are formulated and studied for both
types of systems with and without relaying. The admission control
based joint bandwidth and power allocation problem is formulated
in Section~\ref{sec_ad}, where the greedy search algorithm is also
developed and investigated for both types of systems with and
without relaying. Numerical results are reported in
Section~\ref{sec_sim}, followed by concluding remarks in
Section~\ref{sec_con}.

\section{System Model}\label{sec_md}

\subsection{Without Relaying}
Consider a wireless network, which consists of $M$ source nodes
$S_i$, $i\in \mathcal{M} = \{1,2, \cdots ,M \}$, and $K$
destination nodes $D_i$, $i\in \mathcal{K} = \{1,2, \cdots , K\}$.
The network serves $N$ users $U_i$, $i\in\mathcal{N} =
\{1,2,\cdots,N\}$, where each user represents a one-hop link from
a source to a destination. The set of users which are served by
$S_i$ is denoted by $\mathcal{N}_{S_i}$.

A spectrum of total bandwidth $W$ is available for the
transmission from the sources. This spectrum can be divided into
distinct and nonoverlapping channels of unequal bandwidths, so
that the sources share the available spectrum through frequency
division and, therefore, do not interfere with each other.

Let $P^{\rm S}_i$ and $W^{\rm S}_i$ denote the allocated transmit
power and channel bandwidth of the source to serve $U_i$. Then the
received SNR at the destination of $U_i$ is
\begin{equation}\label{snr}
\gamma^{\rm D}_i = \frac{P^{\rm S}_i h^{\rm SD}_i} {W^{\rm S}_i
N_0}
\end{equation}
where $h^{\rm SD}_i$ denotes the channel gain of the
source--destination link of $U_i$ and $W^{\rm S}_i N_0$ stands for
the power of additive white Gaussian noise (AWGN) over the
bandwidth $W^{\rm S}_i$. The channel gain $h^{\rm SD}_i$ results
from such effects as path loss, shadowing, and fading. Due to the
fact that the power spectral density (PSD) of AWGN is constant
over all frequencies with the constant value denoted by $N_0$, the
noise power in the channel is linearly increasing with the channel
bandwidth. It can be seen from \eqref{snr} that a channel with
larger bandwidth introduces higher noise power and, thus, reduces
the SNR.

Channel capacity gives the upper bound on the achievable rate of a
link. Given $\gamma^{\rm D}_i$, the source--destination link
capacity of $U_i$ is
\begin{equation}\label{cap}
C^{\rm SD}_i = W^{\rm S}_i \log ( 1 + \gamma^{\rm D}_i) = W^{\rm
S}_i \log \left( 1 + \frac{P^{\rm S}_i h^{\rm SD}_i} {W^{\rm S}_i
N_0} \right).
\end{equation}
It can be seen that $W^{\rm S}_i$ characterizes channel bandwidth,
and $\log ( 1 + \gamma^{\rm R}_i)$ characterizes spectral
efficiency and, thus, $C^{\rm SD}_i$ characterizes data rate over
the source--destination link in bits per second. Moreover, for
fixed $W^{\rm S}_i$, $C^{\rm SD}_i$ is a concave increasing
function of $P^{\rm S}_i$. It can be also shown that $C^{\rm
SD}_i$ is a concave increasing function of $W^{\rm S}_i$ for fixed
$P^{\rm S}_i$, although $\gamma^{\rm D}_i$ is a linear decreasing
function of $W^{\rm S}_i$. Indeed, it can be proved that $C^{\rm
SD}_i$ is a concave function of $P^{\rm S}_i$ and $W^{\rm S}_i$
jointly \cite{Kumaran}, \cite{Acharya}, and \cite{Gault}.

\subsection{With Relaying}
Consider $L$ relay nodes $R_i$, $i \in \mathcal{L} = \{1,2,
\cdots, L \}$ added to the network described in the previous
subsection and used to forward the data from the sources to the
destinations. Then each user represents a two-hop link from a
source to a destination via relaying. To reduce the implementation
complexity at the destinations, single relay assignment is adopted
so that each user has one designated relay. Then the set of users
served by $R_i$ is denoted by $\mathcal{N}_{R_i}$. The relays work
in a half-duplex manner due to the practical limitation that they
can not transmit and receive at the same time. A two-phase
decode-and-forward (DF) protocol is assumed, i.e., the relays
receive and decode the transmitted data from the sources in the
first phase, and re-encode and forward the data to the
destinations in the second phase. The sources and relays share the
total available spectrum in the first and second phases,
respectively. It is assumed that the direct links between the
sources and the destinations are blocked and, thus, are not
available. Note that although the two-hop relay model is
considered in the paper, the results are applicable for multi-hop
relay models as well.

Let $P^{\rm R}_i$ and $W^{\rm R}_i$ denote the allocated transmit
power and channel bandwidth of the relay to serve $U_i$. The
two-hop source--destination link capacity of $U_i$ is given by
\begin{equation}\label{capr3}
C^{\rm SD}_i = \min \{C^{\rm SR}_i, C^{\rm RD}_i \} = \left\{
W^{\rm S}_i \log \left( 1 + \frac{P^{\rm S}_i h^{\rm SR}_i}
{W^{\rm S}_i N_0} \right), \, W^{\rm R}_i \log \left( 1 +
\frac{P^{\rm R}_i h^{\rm RD}_i} {W^{\rm R}_i N_0} \right) \right\}
\end{equation}
where $C^{\rm SR}_i$ and $C^{\rm RD}_i$ are the one-hop
source--relay and relay--destination link capacities of $U_i$,
respectively, and $h^{\rm SR}_i$ and $h^{\rm RD}_i$ denote the
corresponding channel gains.

It can be seen from \eqref{capr3} that if equal bandwidth is
allocated to $W^{\rm S}_i$ and $W^{\rm R}_i$, $C^{\rm SR}_i$ and
$C^{\rm RD}_i$ can be unequal due to the power limits on $P^{\rm
S}_i$ and $P^{\rm R}_i$. Then the source--destination link
capacity $C^{\rm SD}_i$ is constrained by the minimum of $C^{\rm
SR}_i$ and $C^{\rm RD}_i$. Note that since all users share the
total bandwidth of the spectrum, equal bandwidth allocation for
all one-hop links can be inefficient. Therefore, the joint
allocation of bandwidth and power is necessary.

\section{Joint Bandwidth and Power Allocation} \label{sec_al}
Different objectives can be considered while jointly allocating
bandwidth and power in wireless multi-user networks. The widely
used objectives for network optimization are (i)~the sum capacity
maximization; (ii)~the worst user capacity maximization; and
(iii)~the total network power minimization. In this section, the
problems of joint bandwidth and power allocation are formulated
for the aforementioned objectives for both considered systems with
and without relaying. It is shown that all these problems are
convex and, therefore, can be efficiently solved using standard
convex optimization methods.

\subsection{Sum Capacity Maximization}
In the applications without delay constraints, a high data rate
from any user in the network is preferable. Thus, it is desirable
to allocate the resources to maximize the overall network
performance, e.g., the sum capacity of all users.

\subsubsection{Without Relaying}
In this case, the joint bandwidth and power allocation problem
aiming at maximizing the sum capacity of all users can be
mathematically formulated as
\begin{subequations}
\begin{eqnarray}
& & \max_{\{P^{\rm S}_i,W^{\rm S}_i\}} \; \sum_{i\in \mathcal{N}}
C^{\rm SD}_i \label{sum1} \\
& & {\rm s.t.}  \sum_{i\in\mathcal{N}_{S_j}} P^{\rm S}_i \leq
P_{S_j}, \ j \in \mathcal{M}\label{sum2} \\
& & \quad \ \ \, \sum_{i\in \mathcal{N}} W^{\rm S}_i
\leq W. \label{sum3}
\end{eqnarray}
\end{subequations}
The nonnegativity constraints on the optimization variables
$\{P^{\rm S}_i,W^{\rm S}_i\}$ are natural and, thus, omitted
throughout the paper for brevity. In the problem
\eqref{sum1}--\eqref{sum3}, the constraint \eqref{sum2} stands
that the total power at $S_j$ is limited by $P_{S_j}$, while the
constraint \eqref{sum3} indicates that the total bandwidth of the
channels allocated to the sources is also limited by $W$.

Note that since $C^{\rm SD}_i$ is a jointly concave function of
$P^{\rm S}_i$ and $W^{\rm S}_i$, the objective function
\eqref{sum1} is convex. The constraints \eqref{sum2} and
\eqref{sum3} are linear and, thus, convex. Therefore, the problem
\eqref{sum1}--\eqref{sum3} itself is convex. Using the convexity,
the closed-form solution of the problem \eqref{sum1}--\eqref{sum3}
is given in the following proposition.

\textbf{Proposition 1:} {\it The optimal solution of the problem
\eqref{sum1}--\eqref{sum3}, denoted by $\{{P^{\rm S}_i}^{\ast},
{W^{\rm S}_i}^{\ast} | i \in \mathcal{N}\}$, is ${P^{\rm
S}_i}^{\ast} = {P^{\rm S}_i}^{\star}$, ${W^{\rm S}_i}^{\ast} =
Wh^{\rm SD}_i {P^{\rm S}_i}^{\star} / \sum_{j \in \mathcal{I}}
h^{\rm SD}_j {P^{\rm S}_j}^{\star}$, $\forall i \in \mathcal{I}$,
and ${P^{\rm S}_i}^{\ast} = {W^{\rm S}_i}^{\ast} = 0$, $\forall i
\notin \mathcal{I}$, where ${P^{\rm S}_i}^{\star} = P_{S_k}$ for
$i \in \mathcal{N}_{S_k}$ and $\mathcal{I} = \{i \, | \, i = {\rm
arg} \max_{j \in \mathcal{N}_{S_k}}h^{\rm SD}_j, k\in \mathcal{M}
\}$.}

\textbf{Proof:} \ See Appendix A.

Proposition~1 shows that for a set of users served by one source,
the sum capacity maximization based allocation strategy allocates
all the power of the source only to the user with the highest
channel gain and, therefore, results in highly unbalanced resource
allocation among the users.

\subsubsection{With Relaying}
The sum capacity maximization based joint bandwidth and power
allocation problem for the network with DF relaying is given by
\begin{subequations}
\begin{eqnarray}
& & \max_{\{P^{\rm S}_i,W^{\rm S}_i,P^{\rm R}_i,W^{\rm R}_i\}} \;
\sum_{i \in \mathcal{N}}C^{\rm SD}_i \label{sumr1} \\
& & {\rm s.t.}  \sum_{i\in\mathcal{N}_{S_j}} P^{\rm S}_i \leq
P_{S_j}, \ j \in \mathcal{M}\label{sumr2} \\
& & \quad \ \, \sum_{i\in\mathcal{N}_{R_j}} P^{\rm R}_i \leq
P_{R_j}, \ j \in \mathcal{L} \label{sumr3} \\
& & \quad \ \ \, \sum_{i\in \mathcal{N}} W^{\rm S}_i \leq W
\label{sumr4} \\
& & \quad \ \ \, \sum_{i\in \mathcal{N}} W^{\rm R}_i \leq W.
\label{sumr5}
\end{eqnarray}
\end{subequations}

Introducing new variables $\{T_i|i \in \mathcal{N}\}$, the problem
\eqref{sumr1}--\eqref{sumr5} can be equivalently rewritten as
\begin{subequations}
\begin{eqnarray}
& & \min_{\{P^{\rm S}_i,W^{\rm S}_i,P^{\rm R}_i,W^{\rm R}_i,T_i\}}
\quad -\sum_{i\in \mathcal{N}}T_i \label{sumrs1} \\
& & {\rm s.t.} \ T_i - C^{\rm SR}_i \leq 0, \ i\in \mathcal{N}
\label{sumrs2} \\
& & \quad \ \ T_i - C^{\rm RD}_i \leq 0, \ i\in \mathcal{N}
\label{sumrs3} \\
& & \quad \ \ \textrm{the \ constraints \ \eqref{sumr2}--\eqref{sumr5}}.
\label{sumrs4}
\end{eqnarray}
\end{subequations}
Note that the constraints \eqref{sumrs2} and \eqref{sumrs3} are
convex since $C^{\rm SR}_i$ and $C^{\rm RD}_i$ are jointly concave
functions of $P^{\rm S}_i$, $W^{\rm S}_i$ and $P^{\rm R}_i$,
$W^{\rm R}_i$, respectively. The constraints \eqref{sumrs4} are
linear and, thus, convex. Therefore, the problem
\eqref{sumrs1}--\eqref{sumrs4} itself is convex. It can be seen
that the closed-form solution of the problem
\eqref{sumrs1}--\eqref{sumrs4} is intractable due to the coupling
of the constraints \eqref{sumrs2} and \eqref{sumrs3}. However, the
convexity of the problem \eqref{sumrs1}--\eqref{sumrs4} allows to
use standard numerical convex optimization algorithms for solving
the problem efficiently \cite{Boyd}.

Intuitively, the sum capacity maximization based allocation for
the network with DF relaying can not result in significantly
unbalanced resource allocation as that for the network without
relaying. It is because the channel gains in both transmission
phases for the networks with relaying affect the achievable
capacity of each user. The following proposition gives the
conditions under which the sum capacity maximization based
resource allocation strategy for the network with relaying does
not allocate any resources to some users.

\textbf{Proposition~2:} {\it If $h^{\rm SR}_i \geq h^{\rm SR}_j$
and $h^{\rm RD}_i \geq h^{\rm RD}_j$ where $\{i,j\} \subseteq
\mathcal{N}_{S_k}$ and $\{i,j\} \subseteq \mathcal{N}_{R_l}$, then
${P^{\rm S}_j}^{\ast} = {W^{\rm S}_j}^{\ast} = {P^{\rm
R}_j}^{\ast} = {W^{\rm R}_j}^{\ast} = 0$.}

\textbf{Proof:} \ See Appendix A.

In particular, if two users are served by the same source and the
same relay, and one user has lower channel gains than the other
user in both transmission phases, then no resource is allocated to
the former user.

\subsection{Worst User Capacity Maximization}
Fairness among users is also an important issue for resource
allocation. If the fairness issue is considered, the achievable
rate of the worst user is commonly used as the network performance
measure. In this case, the joint bandwidth and power allocation
problem for the network without relaying can be mathematically
formulated as
\begin{subequations}
\begin{eqnarray}
& & \max_{\{P^{\rm S}_i,W^{\rm S}_i\}} \quad \min_{i\in
\mathcal{N}} C^{\rm SD}_i \label{mm1}\\
& & {\rm s.t.} \ \textrm{the \ constraints \
\eqref{sum2}--\eqref{sum3} } . \label{mm2}
\end{eqnarray}
\end{subequations}
Similar, for the networks with relaying, the joint bandwidth and
power allocation problem can be formulated as
\begin{subequations}
\begin{eqnarray}
& & \max_{\{P^{\rm S}_i,W^{\rm S}_i,P^{\rm R}_i,W^{\rm R}_i\}}
\quad \min_{i\in \mathcal{N}} C^{\rm SD}_i \label{mmr1}\\
& & {\rm s.t.} \ \textrm{the \ constraints \
\eqref{sumr2}--\eqref{sumr5} } . \label{mmr2}
\end{eqnarray}
\end{subequations}
Introducing a variable $T$, the problem \eqref{mmr1}--\eqref{mmr2}
can be equivalently written as
\begin{subequations}
\begin{eqnarray}
& & \min_{\{P^{\rm S}_i,W^{\rm S}_i,P^{\rm R}_i,W^{\rm R}_i,T\}} \quad - T \label{mmrs1} \\
& & {\rm s.t.} \ T - C^{\rm SR}_i \leq 0, \ i\in \mathcal{N} \label{mmrs2} \\
& & \quad \ \ T - C^{\rm RD}_i \leq 0, \ i\in \mathcal{N} \label{mmrs3} \\
& & \quad \ \ \textrm{the \ constraints \ \eqref{sumr2}--\eqref{sumr5}}. \label{mmrs4}
\end{eqnarray}
\end{subequations}
Similar to the sum capacity maximization based allocation
problems, it can be shown that the problems
\eqref{mm1}--\eqref{mm2} and \eqref{mmrs1}--\eqref{mmrs4} are
convex. Therefore, the optimal solutions can be efficiently
obtained using standard convex optimization methods.

The next proposition indicates that the worst user capacity
maximization based allocation leads to absolute fairness among
users, just the opposite to the sum capacity maximization based
allocation. The proof is intuitive from the fact that the total
bandwidth is shared by all users, and is omitted for brevity.

\textbf{Proposition~3:} {\it In the problems
\eqref{mm1}--\eqref{mm2} and \eqref{mmr1}--\eqref{mmr2}, the
capacities of all users are equal at optimality.}

\subsection{Total Network Power Minimization}
Another widely considered design objective is the minimization of
the total power consumption of all users. This minimization is
performed under the constraint that the rate requirements of all
users are satisfied. The corresponding joint bandwidth and power
allocation problem for the network without relaying can be written
as
\begin{subequations}
\begin{eqnarray}
& & \min_{\{P^{\rm S}_i,W^{\rm S}_i\}} \quad \sum_{i\in
\mathcal{N}} P^{\rm S}_i \label{pm1} \\
& & {\rm s.t.} \ c_i - C^{\rm SD}_i \leq 0, \; i\in \mathcal{N}
\label{pm2} \\
& & \quad \ \ \textrm{the \ constraints \
\eqref{sum2}--\eqref{sum3}} \label{pm3}
\end{eqnarray}
\end{subequations}
while the same problem for the network with relaying is
\begin{subequations}
\begin{eqnarray}
& & \min_{\{P^{\rm S}_i,W^{\rm S}_i,P^{\rm R}_i,W^{\rm R}_i\}}
\quad \sum_{i\in \mathcal{N}} (P^{\rm S}_i + P^{\rm R}_i)
\label{pmr1} \\
& & {\rm s.t.} \ c_i - C^{\rm SR}_i \leq 0, \; i\in \mathcal{N}
\label{pmr2} \\
& & \quad \ \ c_i - C^{\rm RD}_i \leq 0, \; i \in \mathcal{N}
\label{pmr3} \\
& & \quad \ \ \textrm{the \ constraints \
\eqref{sumr2}--\eqref{sumr5}} \label{pmr4}
\end{eqnarray}
\end{subequations}
where $c_i$ is the minimum acceptable capacity for $U_i$ and the
constraints \eqref{pm2} and \eqref{pm3} indicate that the one-hop
link capacities of $U_i$ should be no less than the given capacity
threshold. Similar to the sum capacity maximization and worst user
capacity maximization based allocation problems, the problems
\eqref{pm1}--\eqref{pm3} and \eqref{pmr1}--\eqref{pmr4} are convex
and, thus, can be solved efficiently as mentioned before.

\section{Admission Control Based Joint Bandwidth and Power Allocation}
\label{sec_ad} In the multi-user networks under consideration,
admission control is required if a certain minimum capacity must
be guaranteed for each user. The admission control aims at
maximizing the number of admitted users and is considered next for
both systems with and without relaying.

\subsection{Without Relaying}
The admission control based joint bandwidth and power allocation
problem in the network without relaying can be mathematically
expressed as
\begin{subequations}
\begin{eqnarray}
& & \max_{\{P^{\rm S}_i,W^{\rm S}_i\}, \mathcal{I} \subseteq
\mathcal{N}} \quad |\mathcal{I}| \label{ad1} \\
& & \textrm{s.t.} \ c_i - C^{\rm SD}_i \leq 0,
\ i\in \mathcal{I} \label{ad2} \\
& & \quad \ \ \textrm{the \ constraint \
\eqref{sum2}--\eqref{sum3}} \label{ad3}
\end{eqnarray}
\end{subequations}
where $| \mathcal{I} |$ stands for the cardinality of
$\mathcal{I}$.

Note that the problem \eqref{ad1}--\eqref{ad3} can be solved using
exhaustive search among all possible subsets of users. However,
the computational complexity of the exhaustive search can be very
high since the number of possible subsets of users is
exponentially increasing with the number of users, which is not
acceptable for practical implementation. Therefore, we develop a
suboptimal greedy search algorithm that significantly reduces the
complexity of finding the maximum number of admissible users.

\subsubsection{Greedy Search Algorithm}
Consider the following problem
\begin{subequations}
\begin{eqnarray}
& & G(\mathcal{I}) \ \triangleq \ \min_{\{P^{\rm S}_i,
W^{\rm S}_i\}} \quad \sum_{i \in \mathcal{I}} W^{\rm S}_i
\label{G1} \\
& & \textrm{s.t.} \ c_i - C^{\rm SD}_i \leq 0, \
i \in \mathcal{I} \label{G2} \\
& & \quad \ \ \textrm{the \ constraint \eqref{sum2}} \label{G3}.
\end{eqnarray}
\end{subequations}
The following proposition provides a necessary and sufficient
condition for the admissibility of a set of users.

\textbf{Proposition~4:} {\it A set of users $\mathcal{I}$ is
admissible if and only if $G(\mathcal{I}) \leq W$.}

\textbf{Proof:} \ See Appendix B.

According to Proposition~4, the optimal value $G(\mathcal{I})$ is
the minimum total bandwidth required to support the users in
$\mathcal{I}$, given that all power constraints are satisfied.
This is instrumental in establishing our greedy search algorithm,
which removes users one by one until the remaining users are
admissible. The `worst' user, i.e., the user whose removal reduces
the total bandwidth requirement to the maximum extent, is removed
at each greedy search iteration. In other words, the removal of
the `worst' user results in the minimum total bandwidth
requirement of the remaining users.\footnote{Note that the
approach based on the user removals appears in different contexts
also in \cite{Le}, \cite{Andersin}.} Thus, the removal criterion
can be stated as
\begin{equation} \label{rem}
n {(t)} \triangleq \textrm{arg}\max_{n \in \mathcal{N} {(t-1) }}
\left(G(\mathcal{N} {(t-1)}) - G(\mathcal{N} {(t-1)} \setminus
\{n\}) \right) = \textrm{arg}\min_{n \in \mathcal{N} {(t-1) }}
G(\mathcal{N} {(t-1)} \setminus \{n\})
\end{equation}
where $n {(t)}$ denotes the user removed at the $t$-th greedy
search iteration, $\mathcal{N} {(t)} \triangleq \mathcal{N}
{(t-1)} \setminus \{n {(t)}\}$ denotes the set of remaining users
after $t$ greedy search iterations, and the symbol `$\setminus$'
stands for the set difference operator.

Note that, intuitively, $\mathcal{N} {(t)}$ should be the `best'
set of $N-t$ users that requires the minimum total bandwidth among
all possible sets of $N-t$ users from $\mathcal{N}$, and
$G(\mathcal{N} {(t)})$ is the corresponding minimum total
bandwidth requirement. Thus, the stopping rule for the greedy
search iterations should be finding such $t^{\ast}$ that
$G(\mathcal{N} {(t^{\ast}-1)}) > W$ and $G(\mathcal{N}
{(t^{\ast})})\leq W$. In other words, $N - t^{\ast}$ is expected
to be the maximum number of admissible users, i.e., the optimal
value of the problem \eqref{ad1}--\eqref{ad3}, denoted by
$d^{\ast}$.

\subsubsection{Complexity of the Greedy Search Algorithm}
It can be seen from Proposition~4 that using the exhaustive search
for finding the maximum number of admissible users is equivalent
to checking $G(\mathcal{I})$ for all possible $\mathcal{I}
\subseteq \mathcal{N}$ and, therefore, the number of times of
solving the problem \eqref{G1}--\eqref{G3} is upper bounded by
$\sum^N_{i = d^{\ast}} {N \choose i}$. On the other hand, it can
be seen from \eqref{rem} that using the greedy search, the number
of times of solving the problem \eqref{G1}--\eqref{G3} is upper
bounded by $\sum^{t^{\ast} - 1}_{i = 0} N - i$. Therefore, the
complexity of the proposed greedy search is significantly reduced
as compared to that of the exhaustive search, especially if $N$ is
large and $d^{\ast}$ is small. Moreover, the complexity of the
greedy search can be further reduced. The lemma given below
follows directly from the decomposable structure of the problem
\eqref{G1}--\eqref{G3}, that is, $G(\mathcal{I}) = \sum_{i \in
\mathcal{M}} G(\mathcal{I} \cap \mathcal{N}_{S_i})$.

\textbf{Lemma~1:} {\it The reduction of the total bandwidth
requirement after removing a certain user is only coupled with the
users served by the same source as this user, and is decoupled
with the users served by other sources. Mathematically, it means
that} $G(\mathcal{I}) - G(\mathcal{I} \setminus \{n\}) =
G(\mathcal{I} \cap \mathcal{N}_{S_i}) - G(\mathcal{I} \cap
\mathcal{N}_{S_i} \setminus \{n\})$ \textit{for} $n \in
\mathcal{N}_{S_i}$, $\forall \mathcal{I} \subseteq \mathcal{N}$.

For the sake of brevity, the proof of the lemma is omitted as
trivial. Let $\mathcal{N}_{S_i} {(t)} \triangleq \mathcal{N}_{S_i}
\cap \mathcal{N} {(t)}$ denote the set of remaining users served
by $S_i$ after $t$ greedy search iterations. Applying Lemma~1
directly to the removal criterion in \eqref{rem}, the following
proposition can be obtained.

\textbf{Proposition~5:} {\it The user to be removed at the $t$-th
greedy search iteration according to \eqref{rem} can be found by
first finding the `worst' user in each set of users served by each
source, i.e.,
\[{n^{\ast}_{S_i}} {(t-1)} \triangleq \textrm{arg} \max_{n \in
{\cal N}_{S_i} {(t-1)}} \left(G(\mathcal{N}_{S_i} {(t-1)}) -
G(\mathcal{N}_{S_i} {(t-1)} \setminus \{n\})\right)\] and then
determining the `worst' user among all these 'worst' users.
Mathematically, it means that} $n {(t)} =
{n^{\ast}_{S_{i^{\ast}}}} {(t-1)}$ \textit{where} $i^{\ast}
\triangleq \textrm{arg}\max_{i \in \mathcal{M}}
\left(G(\mathcal{N}_{S_i} {(t-1)}) - G(\mathcal{N}_{S_i} {(t-1)}
\setminus \{{n^{\ast}_{S_i} {(t-1)}}\})\right)$.

The proof is omitted for brevity. Proposition~5 can be directly
used to build an algorithm for searching for the user to be
removed at each greedy search iteration. It is important that such
algorithm has a reduced computational complexity compared to the
direct use of \eqref{rem}. As a result, although the number of
times that the problem \eqref{G1}--\eqref{G3} has to be solved
remains the same, the number of variables of the problem
\eqref{G1}--\eqref{G3} solved at each time is reduced, and is
upper bounded by ${\cal O} ( 8(\max_{i \in {\cal M}} N_{S_i}
)^3)$.

\subsubsection{Optimality Conditions for the Greedy Search
Algorithm} We also study the conditions under which the proposed
greedy search algorithm is optimal. Specifically, the greedy
search is optimal if the set of remaining users after each greedy
search iteration is the `best' set of users, i.e.,
\begin{equation}\label{opt}
\mathcal{N} {(t)} = \mathcal{N}^{\ast}_{N-t}, \forall 1 \leq t
\leq N
\end{equation}
where $\mathcal{N}^{\ast}_{i} \triangleq \textrm{arg} \min_{
|\mathcal{I}| = i} G(\mathcal{I})$ is the 'best' set of $i$ users.

Let us apply the greedy search to the set of users
$\mathcal{N}_{S_i}$ served by the source ${S_i}$. The 'worst'
user, i.e., the user $\bar n_{S_i} {(t)} \triangleq \textrm{arg}
\max_{n \in \bar{\mathcal{N}}_{S_i} {(t-1)}} \left( G(
\bar{\mathcal{N}}_{S_i} {(t-1)}) - G( \bar{\mathcal{N}}_{S_i}
{(t-1)} \setminus \{n\})\right)$ is removed at the $t$-th greedy
search iteration, where $\bar{\mathcal{N}}_{S_i} {(t)} \triangleq
\bar{\mathcal{N}}_{S_i} {(t-1)} \setminus \{\bar{n}_{S_i} {(t)}\}$
denotes the set of remaining users in the set $\mathcal{N}_{S_i}$
after $t$ greedy search iterations. Also let
$\mathcal{N}^{\ast}_{S_i,j} \triangleq \textrm{arg}
\min_{\mathcal{I} \subseteq \mathcal{N}_{S_i}, |\mathcal{I}| = j}
G(\mathcal{I})$ denote the `best' set of $j$ users in
$\mathcal{N}_{S_i}$. The following theorem provides the necessary
and sufficient conditions for the optimality of the proposed
greedy search.

\textbf{Theorem~1:} {\it The condition \eqref{opt} holds if and
only if the following two conditions hold:}

{\it C1: $\bar{\mathcal{N}}_{S_i} {(t)} = \mathcal{N}^{\ast}_{S_i,
N_{S_i} - t}$, $\forall 1 \leq t \leq N_{S_i}$, $\forall i \in
\mathcal{M}$;

C2: $G(\bar{\mathcal{N}}_{S_i} {(t-2)}) - G(
\bar{\mathcal{N}}_{S_i} {(t-1)}) > G( \bar{\mathcal{N}}_{S_i}
{(t-1)}) - G( \bar{\mathcal{N}}_{S_i} {(t)})$, $\forall 2 \leq t
\leq N_{S_i}$, $\forall i \in \mathcal{M}$.}

\textbf{Proof:} \ See Appendix B.

Theorem~1 decouples the optimality condition \eqref{opt} into two
equivalent conditions per each set of users $\mathcal{N}_{S_i}$.
Specifically, the condition C1 indicates that the set of remaining
users in $\mathcal{N}_{S_i}$ after each greedy search iteration is
the `best' set of users, while the condition C2 indicates that the
reduction of the total bandwidth requirement is decreasing with
the greedy search iterations. Theorem~1 allows to focus on
equivalent problems in which users are subject to the same power
constraint.

The following proposition stands that the condition C2 of
Theorem~1 always holds, which reduces the study on the optimality
of the proposed greedy search only to checking the condition C1 of
Theorem~1.

\textbf{Proposition~6:} {\it The condition C2 of Theorem~1 always holds true.}

\textbf{Proof:} \ See Appendix B.

Let $h_i \triangleq h^{\rm SD}_i/N_0$ denote the channel gain
normalized by the noise PSD. Recall that $c_i$ is the minimum
acceptable capacity for $U_i$. Define $F_i(p)$ as the unique
solution for $w$ in the equation
\begin{equation}\label{cap}
c_i = w \log \left( 1 + \frac{h_ip}{w} \right)
\end{equation}
given $h_i$ and $c_i$ for any $p > 0$, which represents the
minimum bandwidth required by a user for its allocated transmit
power. Then the problem \eqref{G1}--\eqref{G3} for the set of
users $\mathcal{N}_{S_i}$ can be rewritten as
\begin{subequations}
\begin{eqnarray}
& & G(\mathcal{N}_{S_i}) \ \triangleq \ \min_{p_i} \quad
\sum_{i\in \mathcal{N}_{S_i}} F_i(p_i)
\label{Gs1} \\
& & \textrm{s.t.} \sum_{i\in \mathcal{N}_{S_i}} p_i \leq P_{S_i} .
\label{Gs2}
\end{eqnarray}
\end{subequations}

The following lemma gives a condition under which C1 holds for a specific $t$.

\textbf{Lemma~2:} {\it If there exists $\mathcal{N}_{S_l,k}
\subseteq \mathcal{N}_{S_l}$, $|\mathcal{N}_{S_l,k}| = k$ such
that $F_i(p) < F_j(p)$, $\forall 0 < p < P_{S_i}$, $\forall i \in
\mathcal{N}_{S_l,k}$ and $\forall j \in \mathcal{N} \setminus
\mathcal{N}_{S_l,k}$, then $\mathcal{N}_{S_l,k} =
\mathcal{N}^{\ast}_{S_l,k} = \bar{\mathcal{N}}_{N_{S_l}} {(N_{S_l}
- k)}$.}

\textbf{Proof:} \ See Appendix B.

It can be seen from Lemma~2 that since any user in
$\mathcal{N}_{S_l,k}$ has a smaller bandwidth requirement than any
user in $\mathcal{N} \setminus \mathcal{N}_{S_l,k}$ for the same
allocated power over the available power range, the former is
preferable to the latter in the sense of reducing the total
bandwidth requirement. Therefore, $\mathcal{N}_{S_l,k}$ is the
`best' set of $k$ users and the greedy search removes users in
$\mathcal{N} \setminus \mathcal{N}_{S_l,k}$ before
$\mathcal{N}_{S_l,k}$.

It is worth noting that C1 does not hold in general. Indeed, since
the reduction of the total bandwidth requirement is maximized only
at each single greedy search iteration, the greedy search does not
guarantee that the reduction of the total bandwidth requirement is
also maximized over multiple greedy search iterations. In other
words, it does not guarantee that the set of remaining users is
the `best' set of users. In order to demonstrate this, we present
the following counter example.

\textbf{Example~1:} {\it Let $\mathcal{N}_{S_1}=\{1, 2, 3\}$. Also
let $h_1=4$, $h_2=5$, $h_3=6$, $c_1=1$, $c_2=1.1$, $c_3=1.2$, and
$P_{S_1}=1.1$. Then we have $G(\{1,2\}) = 1.3849$, $G(\{1,3\}) =
1.3808$, $G(\{2,3\}) = 1.3573$, $G(\{1\}) = 0.4039$, $G(\{2\}) =
0.4135$, $G(\{3\}) = 0.4292$ and, therefore,
$\bar{\mathcal{N}}_{S_1} {(1)} = \{2,3\}$,
$\bar{\mathcal{N}}_{S_1} {(2)} = \{2\}$,
$\mathcal{N}^{\ast}_{S_1,1} = \{1\}$. This shows that
$\bar{\mathcal{N}}_{S_1} {(2)} \neq \mathcal{N}^{\ast}_{S_1,1}$.}

Example~1 shows that the `worst' user, which is removed first in
the greedy search, may be among the `best' set of users after more
users are removed. An intuitive interpretation of this result is
that the bandwidth required by the `worst' user changes from being
larger to being smaller compared to the bandwidth required by
other users for the same allocated power. It is because the
average available power to each user increases after some users
are removed in the greedy search.

Using Lemma~2, the following proposition that gives a sufficient
condition under which C1 holds is in order.

\textbf{Proposition~7:} {\it The condition C1 holds if for any $i
\in \mathcal{N}_{S_k}$, $\forall k \in \mathcal{M}$, there exists
no more than one $j \in \mathcal{N}_{S_k}$, $j \neq i$, such that}

{\it C3: $F_i (p)$ intersects $F_j (p)$ in the interval $0 < p <
P_{S_i}$. }

\textbf{Proof:} \ See Appendix B.

It can be seen from Proposition~7 that the chance that C1 holds
increases when the chance that C3 holds decreases. Moreover, the
chance that C1 holds increases when $N_{S_i}$ is large for all $i
\in \mathcal{M}$ or $M$ is large. The next lemma compares the
bandwidth requirements of two users in terms of the ratio between
their minimum acceptable capacities and the ratio between their
channel gains.

\textbf{Lemma~3:} {\it If $i \neq j$ and $h_j/h_i \geq 1$, then

(i)~$\forall p, \; F_i(p)$ intersects $F_j (p)$ at a unique point
$p^\prime$, if and only if $1 < c_j/c_i < h_j/h_i $; furthermore,
$p^{\prime}$ increases as $h_j/h_i$ increases or $c_j/c_i$
decreases;

(ii)~$F_i(p) > F_j(p)$, $\forall p > 0$, or $F_i(p) = F_j(p)$,
$\forall p > 0$, if and only if $c_j/c_i \leq 1$;

(iii)~$F_i(p) < F_j(p)$, $\forall p > 0$, if and only if $c_j/c_i
\geq h_j/h_i$.}

\textbf{Proof:} \ See Appendix B.

It can be seen from Lemma~3 that the condition C3 of Proposition~7
holds if and only if the claim (i) of Lemma~3 holds with $0 <
p^{\prime} < P_{S_i}$. Then it follows from Proposition~7 and
Lemma~3 that the chance that the condition C1 of Theorem~1 holds
increases for smaller $h_j/h_i$ and larger $c_j/c_i$ or for
smaller $P_{S_i}$. Moreover, when $h_j/h_i$ is infinitely large,
$c_j = c_i$ or when $P_{S_i}$ is infinitely small, the condition
C1 of Theorem~1 always holds. Therefore, the condition C1 of
Theorem~1 is, in fact, a mild condition and it always holds when
the diversity of user rate requirements differs sufficiently from
that of user channel gains, or the available source power is
small, or the number of users served by a source is small, or the
number of sources is small.

Applying Lemma~3, Proposition~6, and Proposition~7, the next
corollary follows directly from Theorem~1.

\textbf{Corollary~1:} {\it The proposed greedy search is optimal,
i.e., $\mathcal{N} {(t)} = \mathcal{N}^{\ast}_{N-t}$, $\forall 1
\leq t \leq N$, if $c_i = c_j$, $\forall i, j \in \mathcal{N}$, $i
\neq j$.}

Note that the optimality condition given in \eqref{opt} is a
sufficient condition under which $\mathcal{N} {(t^{\ast})} =
\mathcal{N}^{\ast}_{N - t^{\ast}} =
\mathcal{N}^{\ast}_{d^{\ast}}$. Indeed, the greedy search is
optimal if and only if $t^{\ast} = N - d^{\ast}$. Therefore, even
if $\mathcal{N} {(t^{\ast})} \neq \mathcal{N}^{\ast}_{d^{\ast}}$,
the greedy search still gives the maximum number of admissible
users if $G(\mathcal{N}^{\ast}_{d^{\ast}}) < G(\mathcal{N} {(N -
d^{\ast})}) \leq W$.

\subsection{With Relaying}

The admission control based joint bandwidth and power allocation
problem in the network with relaying is given by
\begin{subequations}
\begin{eqnarray}
& & \max_{\{P^{\rm S}_i,W^{\rm S}_i,P^{\rm R}_i,W^{\rm R}_i\},
\mathcal{I} \subseteq \mathcal{N}} \quad |\mathcal{I}|
\label{adr1} \\
& & \textrm{s.t.} \ c_i - C^{\rm SR}_i \leq 0, \ i\in
\mathcal{I} \label{adr2} \\
& & \quad \ \ c_i - C^{\rm RD}_i \leq 0, \ i\in
\mathcal{I} \label{adr3}\\
& & \quad \ \ \textrm{the \ constraint \
\eqref{sumr2}--\eqref{sumr5}}. \label{adr4}
\end{eqnarray}
\end{subequations}
The proposed greedy search algorithm can also be used to reduce
the complexity of solving the problem \eqref{adr1}--\eqref{adr4}.
Specifically, the problem \eqref{adr1}--\eqref{adr4} can be
decomposed into
\begin{subequations}
\begin{eqnarray}
& & \max_{\{P^{\rm S}_i,W^{\rm S}_i\}, \mathcal{I} \subseteq
\mathcal{N}} \quad |\mathcal{I}| \label{adr1_1} \\
& & \textrm{s.t.} \ c_i - C^{\rm SR}_i \leq 0, \ i\in
\mathcal{I} \label{adr1_2} \\
& & \quad \ \ \textrm{the \ constraint \ \eqref{sumr2},
\eqref{sumr4}} \label{adr1_3}
\end{eqnarray}
\end{subequations}
and
\begin{subequations}
\begin{eqnarray}
& & \max_{\{P^{\rm R}_i,W^{\rm R}_i\}, \mathcal{I} \subseteq
\mathcal{N}} \quad |\mathcal{I}| \label{adr2_1} \\
& & \textrm{s.t.} \ c_i - C^{\rm RD}_i \leq 0, \ i\in
\mathcal{I} \label{adr2_2} \\
& & \quad \ \ \textrm{the \ constraint \ \eqref{sumr3},
\eqref{sumr5}}. \label{adr2_3}
\end{eqnarray}
\end{subequations}
each of which has the same form as the problem
\eqref{ad1}--\eqref{ad3}. Therefore, the proposed greedy search
can be applied for solving each of these two problems separately.
As a result, the numbers of users removed by the greedy search in
each transmission phase can be found as $t^{\ast}_1$ and
$t^{\ast}_2$, respectively. Let $d^{\ast}$, $d^{\ast}_1$, and
$d^{\ast}_2$ denote the optimal values of the problem
\eqref{adr1}--\eqref{adr4}, \eqref{adr1_1}--\eqref{adr1_3}, and
\eqref{adr2_1}--\eqref{adr2_3}, respectively. Since the feasible
set of the problem \eqref{adr1}--\eqref{adr4} is a subset of those
of the problem \eqref{adr1_1}--\eqref{adr1_3} and
\eqref{adr2_1}--\eqref{adr2_3}, we have $d^{\ast} \leq \min
\{d^{\ast}_1, d^{\ast}_2\}$. Therefore, $d^{\ast}$ should be
obtained by solving the problem
\begin{subequations}
\begin{eqnarray}
& & \max_{\{P^{\rm S}_i,W^{\rm S}_i,P^{\rm R}_i,W^{\rm R}_i\},
\mathcal{I} \subseteq \mathcal{N}, |\mathcal{I}| \leq t^{\prime}}
\quad |\mathcal{I}| \label{adr3_1} \\
& & \textrm{s.t.} \ c_i - C^{\rm SR}_i \leq 0, \ i\in
\mathcal{I} \label{adr3_2} \\
& & \quad \ \ c_i - C^{\rm RD}_i \leq 0, \ i\in \mathcal{I}
\label{adr3_3}\\
& & \quad \ \ \textrm{the \ constraints \
\eqref{sumr2}--\eqref{sumr5}} \label{adr3_4}
\end{eqnarray}
\end{subequations}
where $d^{\prime} \triangleq \min \{N - t^{\ast}_1, N -
t^{\ast}_2\}$ and the feasible set is reduced as compared to that
of the problem \eqref{adr1}--\eqref{adr4}. The problem
\eqref{adr3_1}--\eqref{adr3_4} can then be solved using exhaustive
search with significantly reduced complexity compared to total
exhaustive search over two transmission phases.

Using the exhaustive search, the number of times that the problem
\eqref{G1}--\eqref{G3} has to be solved is upper bounded by $2
\sum^N_{i = d^{\ast}} {N \choose i}$. Using the greedy search
combined with the exhaustive search, this number of times
significantly reduces and is upper bounded by $\sum^{t^{\ast}_1 -
1}_{i = 0} N - i + \sum^{t^{\ast}_2 - 1}_{i = 0} N - i + 2
\sum^{d^{\prime}}_{i = d^{\ast}} {N \choose i}$ if $d^{\prime}
\geq d^{\ast}$ and $\sum^{t^{\ast}_1 - 1}_{i = 0} N - i +
\sum^{t^{\ast}_2 - 1}_{i = 0} N - i + 2 {N \choose d^{\prime}}$ if
$d^{\prime} < d^{\ast}$. This complexity reduction is especially
pronounced when $N$ is large and $d^{\prime}$, $d^{\ast}$ are
small.

It can be seen from comparing the problem
\eqref{adr1}--\eqref{adr4} and \eqref{adr3_1}--\eqref{adr3_4} that
the greedy search is optimal if and only if $d^{\prime} \geq
d^{\ast}$.

\section{Simulation Results}\label{sec_sim}

\subsection{Joint Bandwidth and Power Allocation}

Consider a wireless network which consists of four users
$\mathcal{N} = \{1,2,3,4\}$, four sources, and two relays. The
source and relay assignments to the users are the following:
$\mathcal{N}_{S_1}=\{1\}$, $\mathcal{N}_{S_2}=\{2\}$,
$\mathcal{N}_{S_3}=\{3\}$, $\mathcal{N}_{S_4}=\{4\}$,
$\mathcal{N}_{R_1}=\{1,2\}$, and $\mathcal{N}_{R_2}=\{3,4\}$. The
sources and destinations are randomly distributed inside a square
area bounded by (0,0) and (10,10), and the relays are fixed at
(5,3) and (5,7). The path loss and the Rayleigh fading effects are
present in all links. The path loss gain is given by $g =
(1/d)^2$, where $d$ is the distance between two transmission ends,
and the variance of the Rayleigh fading gain is denoted as
$\sigma^2$. We set $P_{S_i}=20$, $\forall i\in \{1,2,3,4\}$,
$P_{R_i} \triangleq P_R = 40$, $\forall i \in \{1,2\}$, $W = 10$,
$\sigma^2=5$, and $c_i = 1$, $\forall i\in \{1,2,3,4\}$ as default
values if no other values are indicated otherwise. The noise PSD
$N_0$ equals to 1. All results are averaged over 1000 simulation
runs for different instances of random channel realizations.

The following resource allocation schemes are compared to each
other: the proposed optimal joint bandwidth and power allocation
(OBPA), equal bandwidth with optimal power allocation (EBOPA), and
equal bandwidth and power allocation (EBPA). Software package
TOMLAB \cite{Tomlab} is used to solve the corresponding convex
optimization problems.

\begin{figure}
\subfigure[]{\label{sc_pr}\begin{minipage}[b]{1.0\linewidth}
  \centering
  \centerline{\epsfig{figure=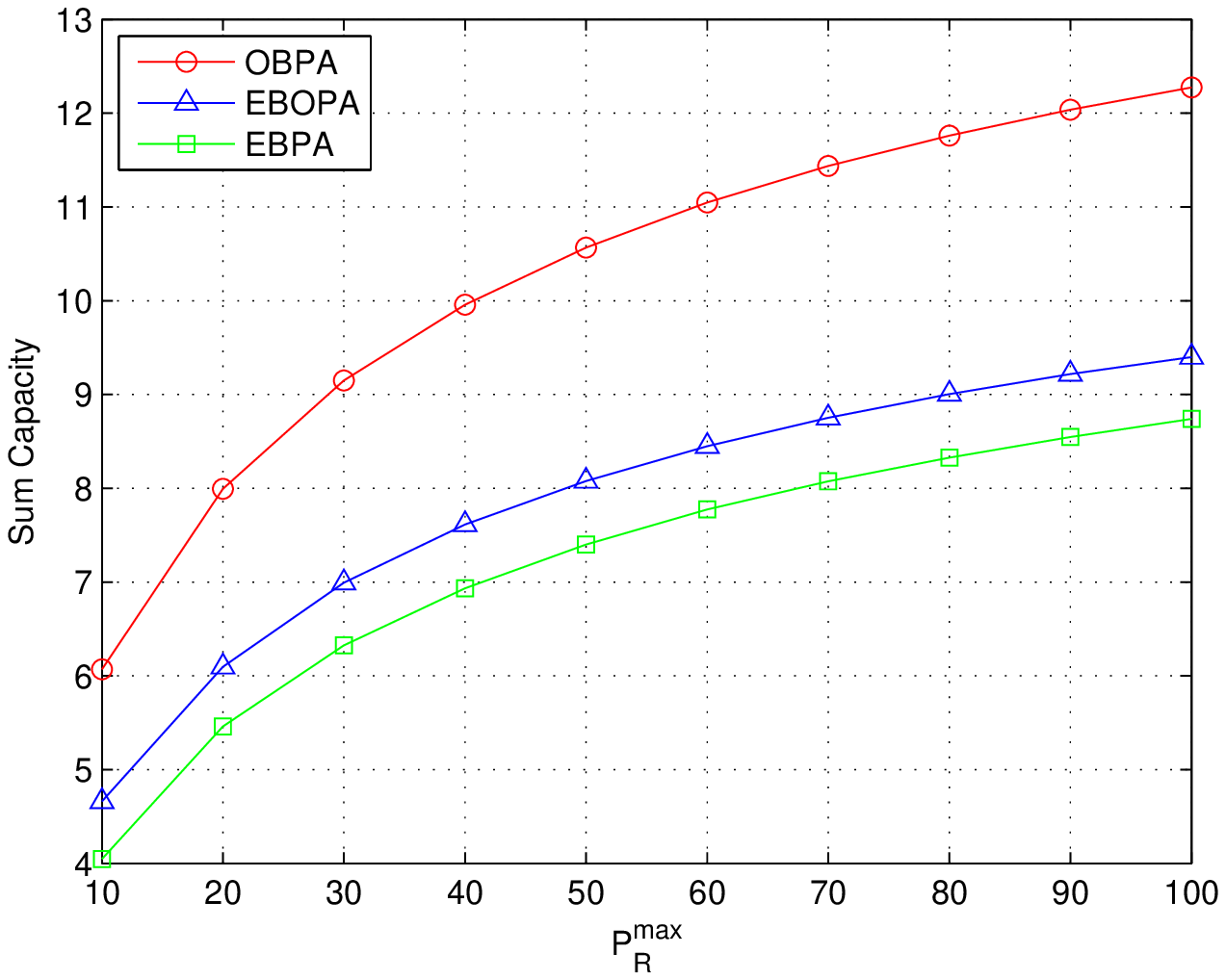,width=8.2cm}}
\end{minipage}}
\subfigure[]{\label{sc_w}\begin{minipage}[b]{1.0\linewidth}
  \centering
  \centerline{\epsfig{figure=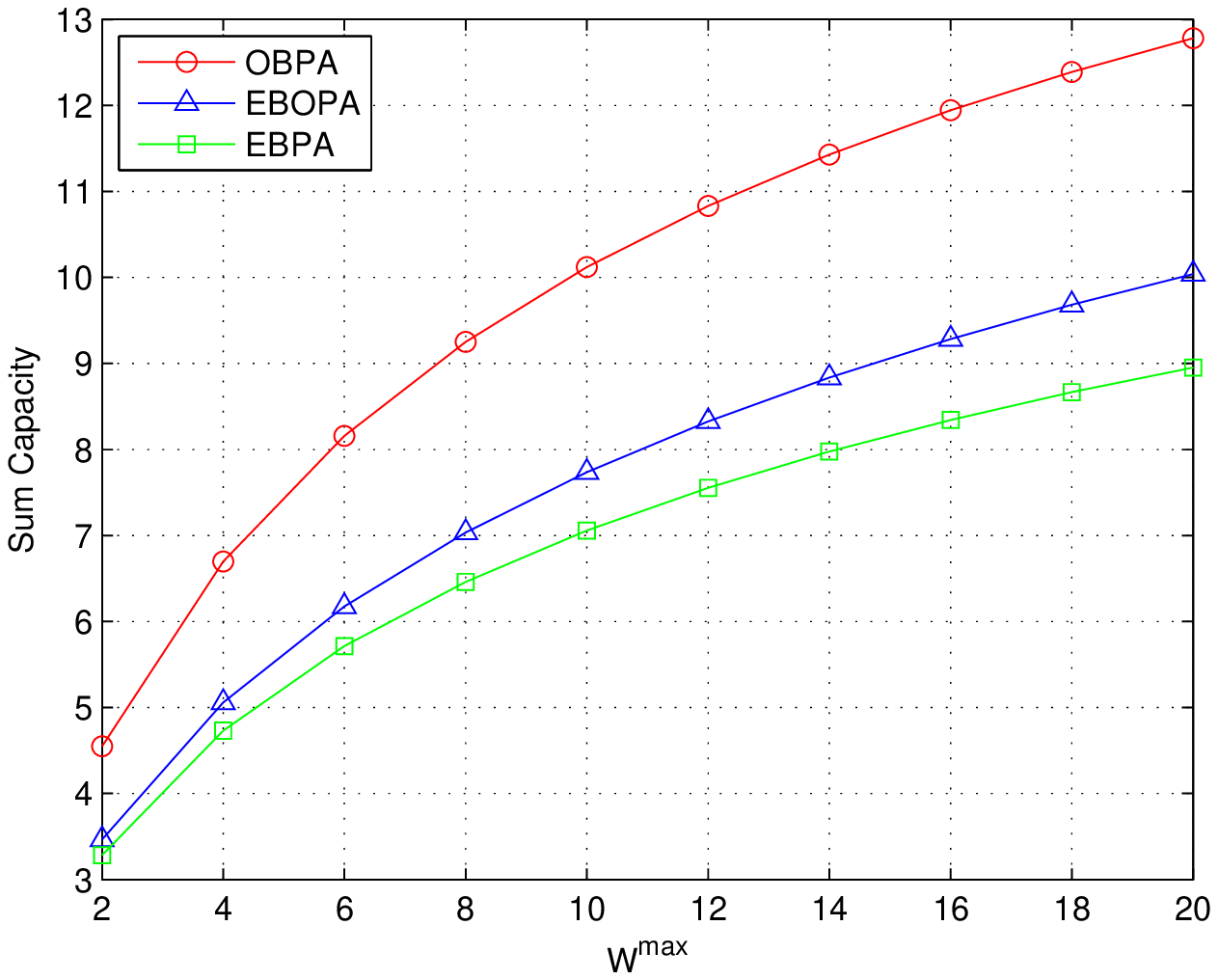,width=8.2cm}}
\end{minipage}}
\caption{Sum capacity maximization based allocation: (a) $W = 10$,
(b) $P_R = 40$.} \label{fg_sc}
\end{figure}

In Figs.~\ref{fg_sc}~(a)~and~(b), the performance of the sum
capacity maximization based allocation is shown versus $P_R$ and
$W$, respectively. These figures show that the OBPA scheme
achieves about 30\% to 50\% performance improvement over the other
two schemes for all parameter values. The performance improvement
is higher when $P_R$ or $W$ is larger. The observed significant
performance improvement for the OBPA can be partly attributed to
the fact that the sum capacity maximization based joint bandwidth
and power allocation can lead to highly unbalanced resource
allocation, while bandwidth is equally allocated in the EBOPA and
both bandwidth and power are equally allocated in the EBPA.

\begin{figure}
\subfigure[]{\label{mm_pr}\begin{minipage}[b]{1.0\linewidth}
  \centering
  \centerline{\epsfig{figure=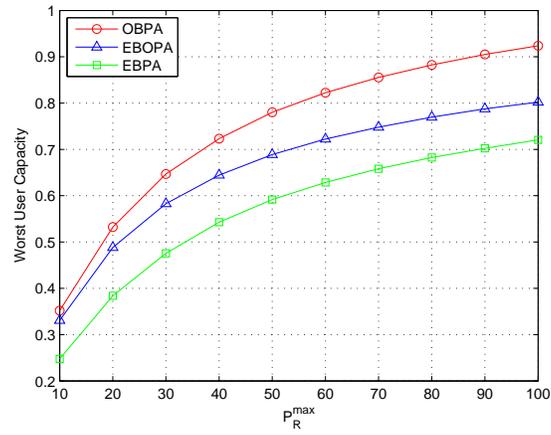,width=8.2cm}}
\end{minipage}}
\subfigure[]{\label{mm_w}\begin{minipage}[b]{1.0\linewidth}
  \centering
  \centerline{\epsfig{figure=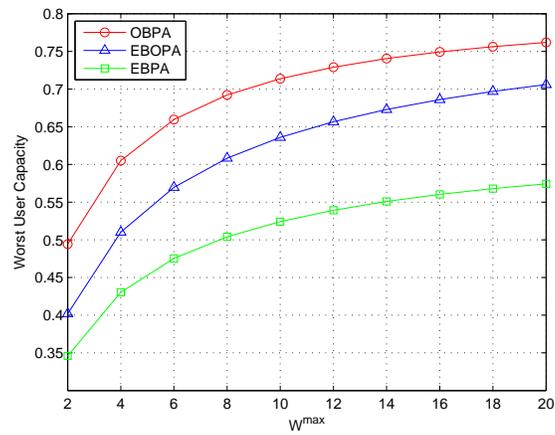,width=8.2cm}}
\end{minipage}}
\caption{Worst user capacity maximization based allocation: (a) $W
= 10$, (b) $P_R = 40$.} \label{fg_mm}
\end{figure}

Figs.~\ref{fg_mm}~(a)~and~(b) demonstrate the performance of the
worst user capacity maximization based allocation versus $P_R$ and
$W$, respectively. The performance improvement for the OBPA is
about 10\% to 30\% as compared to the EBOPA for all parameter
values. The improvement provided by the OBPA, in this case, is not
as significant as that in Figs.~\ref{fg_sc}~(a)~and~(b), which can
be attributed to the fact that the worst user capacity
maximization based allocation results in relatively balanced
resource allocation, while the EBOPA and the EBPA are balanced
bandwidth and totally balanced allocation schemes, respectively.

\begin{figure}
\subfigure[]{\label{pm_th}\begin{minipage}[b]{1.0\linewidth}
  \centering
  \centerline{\epsfig{figure=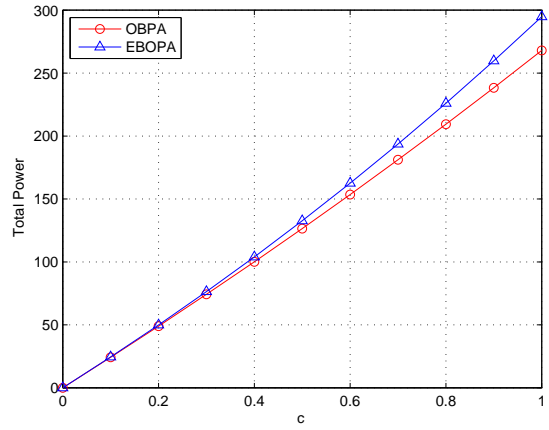,width=8.2cm}}
\end{minipage}}
\subfigure[]{\label{pm_w}\begin{minipage}[b]{1.0\linewidth}
  \centering
  \centerline{\epsfig{figure=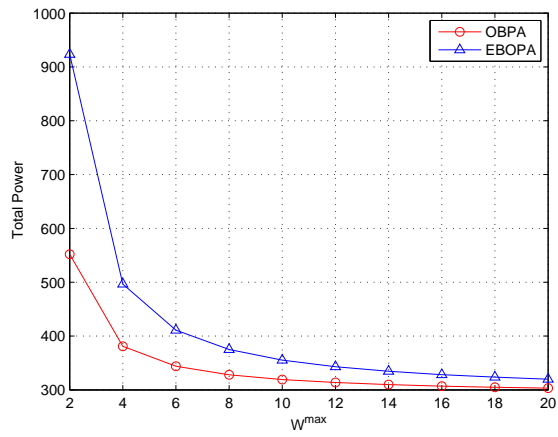,width=8.2cm}}
\end{minipage}}
\caption{Power minimization based allocation: (a) $W = 10$, (b) $c
= 1$.} \label{fg_pm}
\end{figure}

Figs.~\ref{fg_pm}~(a)~and~(b) show the total power consumption of
the sources and relays versus $c$ and $W$ for the power
minimization based allocation, where $c_1 = c_2 = c_3 = c_4
\triangleq c$ is assumed. Note that the total power of the OBPA is
always about 10\% to 30\% less than that of the EBOPA, and the
total power difference between the two tested schemes is larger
when $c$ is larger, or when $W$ is smaller. This shows that more
power is saved when the parameters are unfavorable due to the
flexible bandwidth allocation in the OBPA.

\begin{figure}[t]
\begin{minipage}[b]{1.0\linewidth}
  \centering
  \centerline{\epsfig{figure=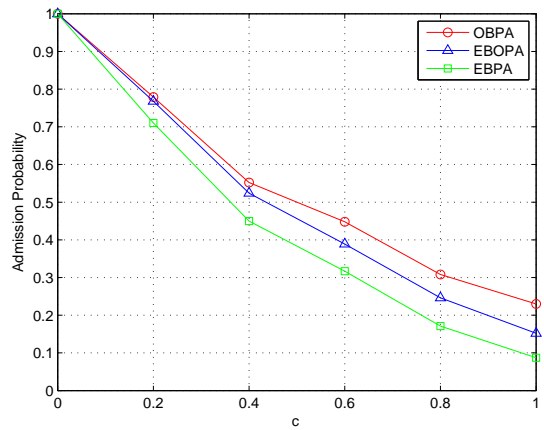,width=8.2cm}}
\end{minipage}
\caption{Admission probability vs capacity threshold.}\label{fg_ad}
\end{figure}

Fig.~\ref{fg_ad} depicts the admission probability versus $c$,
where $c_1 = c_2 = c_3 = c_4 \triangleq c$ is assumed. The
admission probability is defined as the probability that $c$ can
be satisfied for all the users under random channel realizations.
The figure shows that the OBPA outperforms the other two schemes
for all values of $c$, and the improvement is more significant
when $c$ is large. This shows that more users or users with higher
rate requirements can be admitted into the network using the OBPA
scheme.

\subsection{Greedy Search Algorithm}

In this example, the performance of the proposed greedy search
algorithm is compared to that of the exhaustive search algorithm.
We consider eight users $\mathcal{N} = \{1,2,\cdots,8\}$
requesting for admission. The sources and the destinations are
randomly distributed inside a square area bounded by (0,0) and
(10,10). We assume that $c_i$, $i \in \{1,2,\cdots,8\}$, is
uniformly distributed over the interval $[c_0,c_0 + 4]$ where
$c_0$ is a variable parameter. The channel model is the same as
that given in the previous subsection. We set $W = 10$, $\sigma^2
= 10$ as default values. The results are averaged over 20 random
channel realizations.

\subsubsection{Without Relaying}

We consider the following two network setups.

\begin{figure}
\subfigure[setup
1]{\label{fg_gdy1}\begin{minipage}[b]{1.0\linewidth}
  \centering
  \centerline{\epsfig{figure=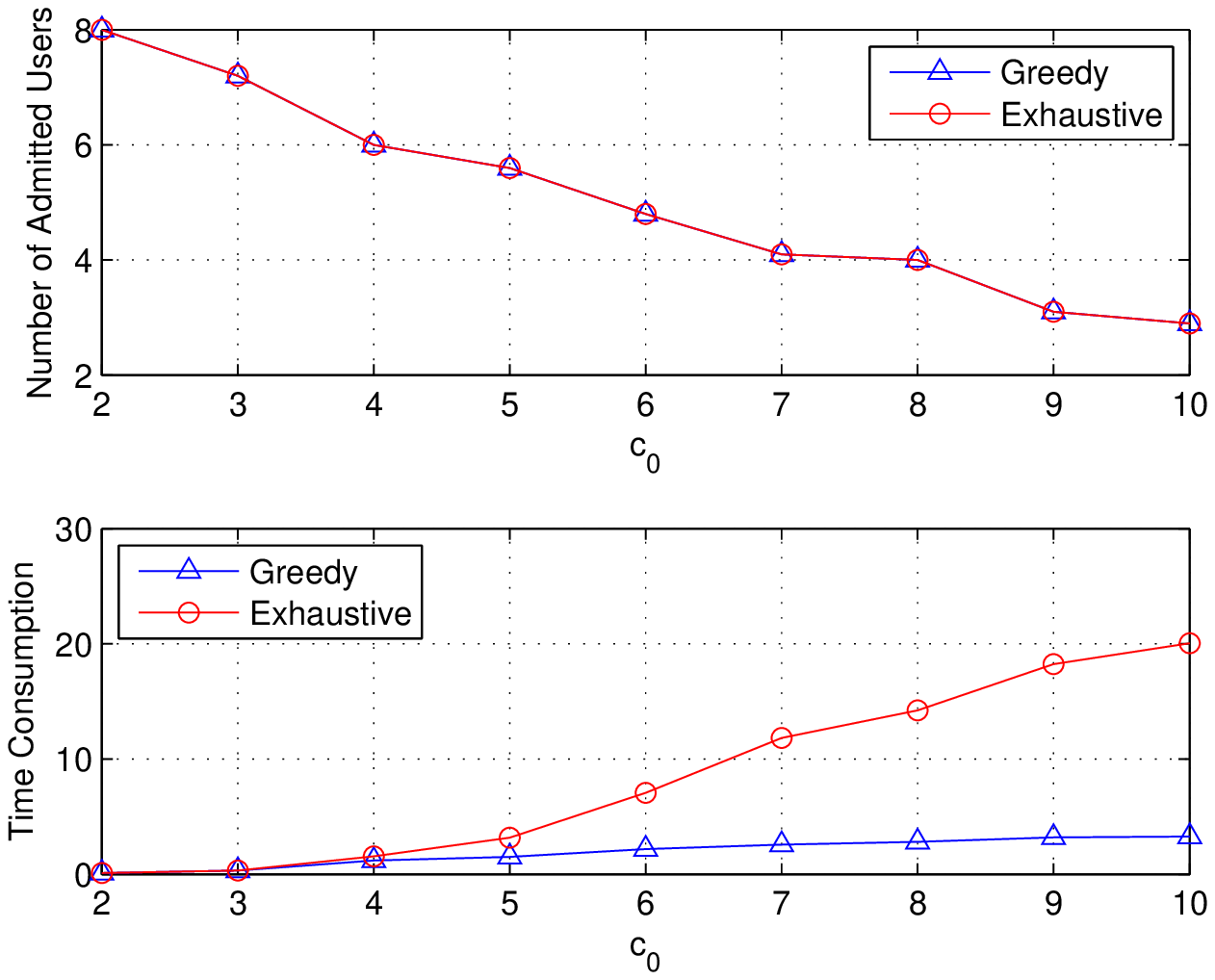,width=8.2cm}}
\end{minipage}}
\subfigure[setup
2]{\label{fg_gdy2}\begin{minipage}[b]{1.0\linewidth}
  \centering
  \centerline{\epsfig{figure=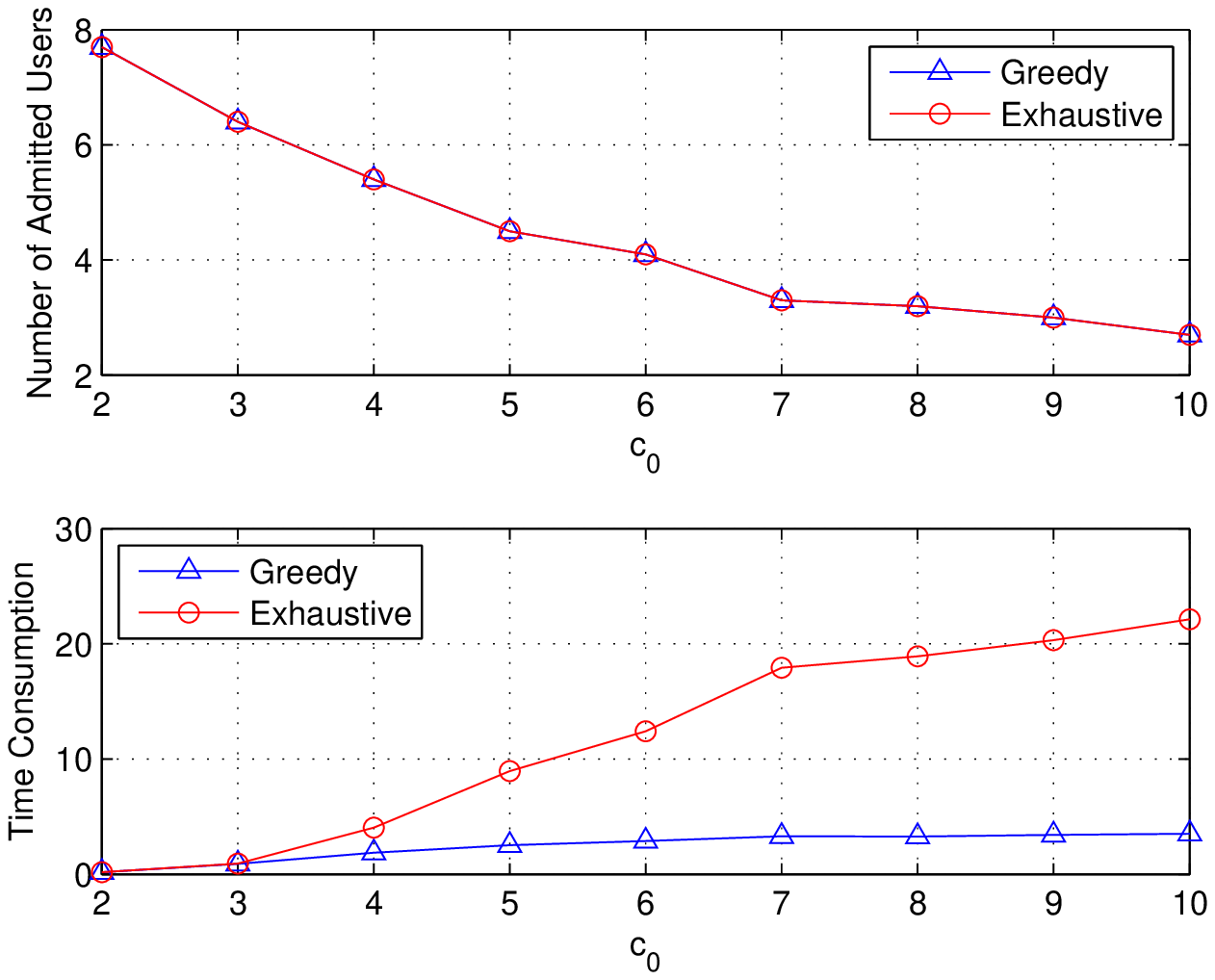,width=8.2cm}}
\end{minipage}}
\caption{Greedy search algorithm vs exhaustive search algorithm.}
\end{figure}

\textit{Setup 1:} In this setup, the optimality condition of the
greedy search is satisfied. Specifically, there are four sources.
The source assignments to the users are the following:
$\mathcal{N}_{S_1} = \{1,2\}$, $\mathcal{N}_{S_2} = \{3,4\}$,
$\mathcal{N}_{S_3} = \{5,6\}$, and $\mathcal{N}_{S_4} = \{7,8\}$.
We set $P_{S_i} = 40$, $\forall i \in \{1,2,3,4\}$.
Fig.~\ref{fg_gdy1} shows the number of admitted users obtained by
the greedy search and the corresponding computational complexity
in terms of the running time versus $c_0$. The figure shows that
the greedy search gives exactly the same number of admitted users
as that of the exhaustive search for all values of $c_0$. This
confirms that the optimal solution can be obtained when the
optimality condition of the greedy search is satisfied. The time
consumption of the greedy search is significantly less than that
of the exhaustive search, especially when $c_0$ is large. This
shows that the proposed algorithm is especially efficient when the
number of candidate users is large and the number of admitted
users is small.

\textit{Setup 2:} In this setup, the optimality condition of the
greedy search may not be satisfied. There are two sources and the
source assignments to the users are the following:
$\mathcal{N}_{S_1} = \{1,2,3,4\}$, and $\mathcal{N}_{S_2} =
\{5,6,7,8\}$. We set $P_{S_i} = 80$, $\forall i \in \{1,2\}$.
Fig.~\ref{fg_gdy2} demonstrates the performance of the greedy
search. Similar conclusions can be drawn for this setup as those
for Setup 1. This indicates that the proposed greedy search
algorithm can still perform optimally even if the sufficient
optimality condition may not be satisfied.

\subsubsection{With Relaying}

We also consider two network setups as follows.

\begin{figure}
\subfigure[setup
3]{\label{fg_gdy_r1}\begin{minipage}[b]{1.0\linewidth}
  \centering
  \centerline{\epsfig{figure=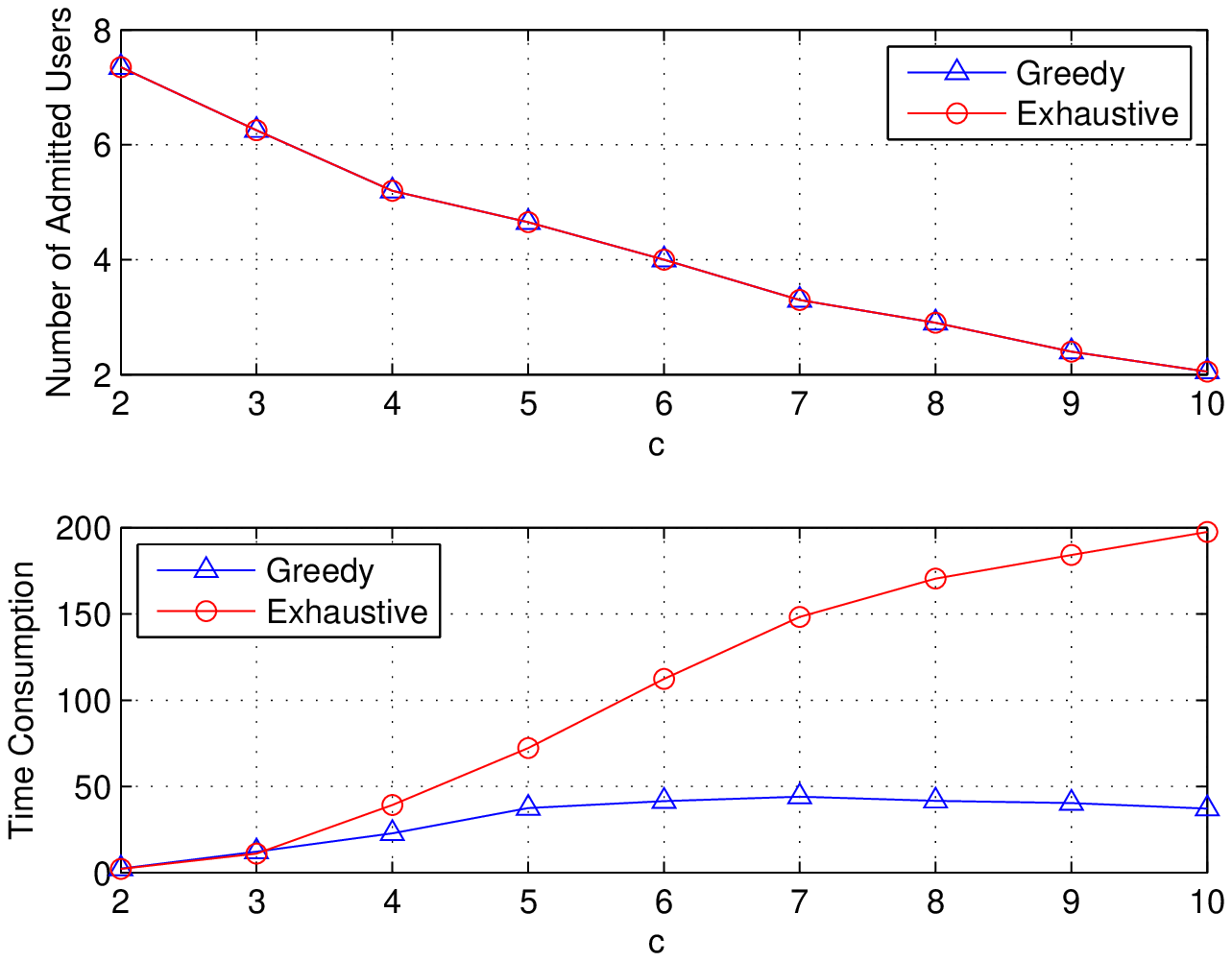,width=8.2cm}}
\end{minipage}}
\subfigure[setup
4]{\label{fg_gdy_r2}\begin{minipage}[b]{1.0\linewidth}
  \centering
  \centerline{\epsfig{figure=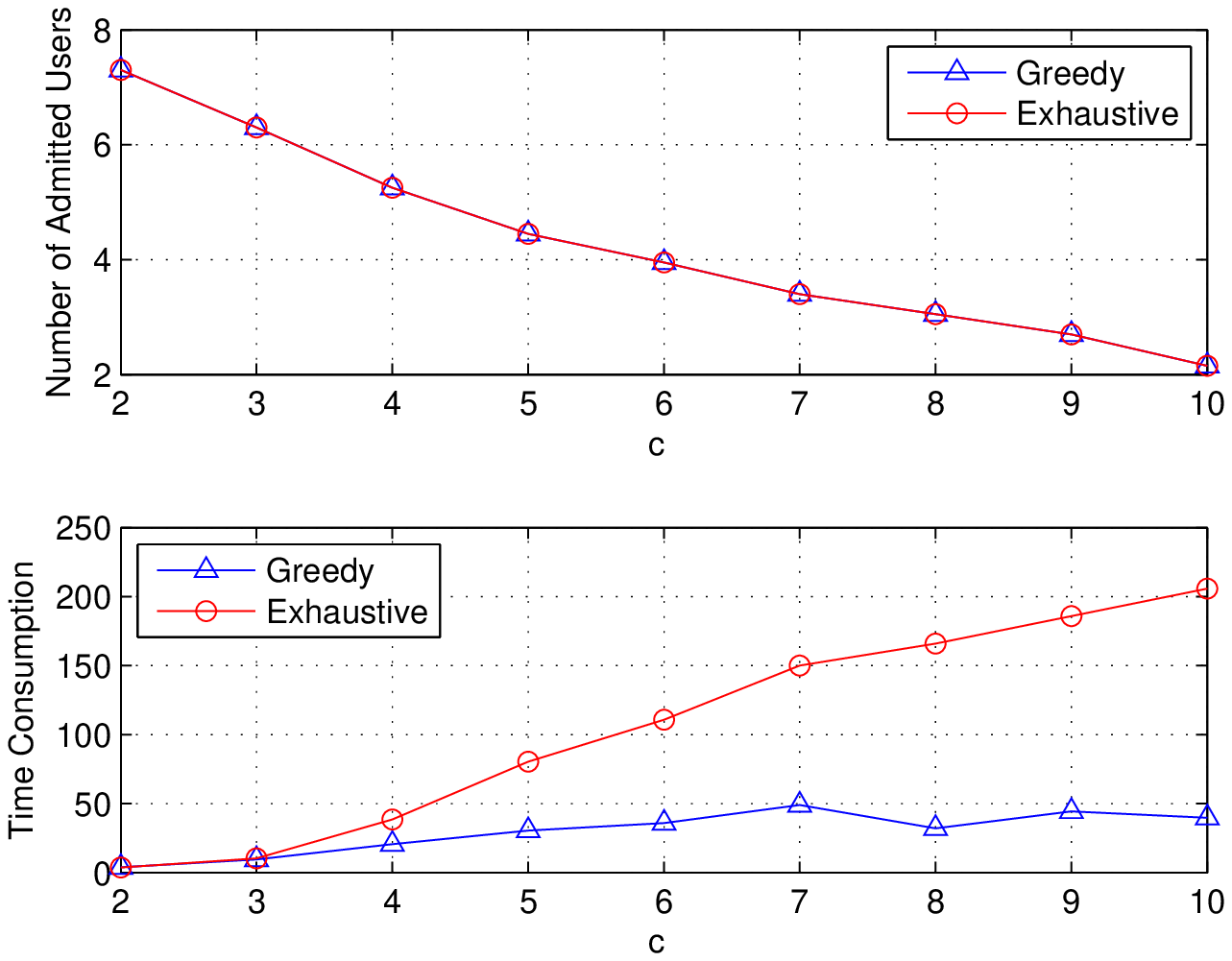,width=8.2cm}}
\end{minipage}}
\caption{Greedy search algorithm vs exhaustive search algorithm.}
\end{figure}

\textit{Setup 3:} In this setup, the optimality condition of the
greedy search is satisfied. Specifically, in addition to the Setup
1 given in the case without relaying, four relays are included
with the following user assignments: $\mathcal{N}_{R_1} =
\{1,2\}$, $\mathcal{N}_{R_2} = \{3,4\}$,
$\mathcal{N}_{R_3}=\{5,6\}$, and $\mathcal{N}_{R_4}=\{7,8\}$. The
relays are fixed at (5,2), (5,4), (5,6), and (5,8), and we set
$P_{R_i} = 40$, $\forall i \in \{1,2,3,4\}$. Fig.~\ref{fg_gdy_r1}
shows the number of admitted users obtained by the greedy search
and the corresponding computational complexity in terms of the
running time versus $c_0$. Similar observations can be obtained as
those for Setup 1. However, it can be noted as expected that the
time consumption of the greedy search for the network with
relaying is much more than that for the network without relaying.

\textit{Setup 4:} In this setup, the optimality condition of the
greedy search may not be satisfied. Specifically, in addition to
the Setup 2 given in the case without relaying, two relays are
included with the following user assignments: $\mathcal{N}_{R_1} =
\{1,2,7,8\}$, $\mathcal{N}_{R_2} = \{3,4,5,6\}$. The relays are
fixed at (5,3) and (5,7) and we also set $P_{R_i} = 80$, $\forall
i\in \{1,2\}$. Fig.~\ref{fg_gdy_r2} demonstrates the performance
of the greedy search. Similar conclusions can be obtained as those
for Setup 3.

\section{Conclusion}\label{sec_con}
In this paper, joint bandwidth and power allocation has been proposed for wireless multi-user networks with and without
relaying to (i)~maximize the sum capacity of all users; (ii)~maximize the capacity of the worst user; (iii)~minimize the total
power consumption of all users. It is shown that the corresponding resource allocation problems are convex and, thus, can be
solved efficiently. Moreover, admission control based joint bandwidth and power allocation has been considered. Because of the
high complexity of the admission control problem, a suboptimal greedy search algorithm with significantly reduced complexity
has been developed. The optimality condition of the proposed greedy search has been derived and analyzed. Simulation results
demonstrate the efficiency of the proposed allocation schemes and the advantages of the greedy search.

\section*{Appendix A: Proofs of Lemmas, Propositions, and Theorems in Section~\ref{sec_al} }

\textit{Proof of Proposition~1:} We first give the following lemma.

\textbf{Lemma 4:} {\it The optimal solution of the problem
\begin{subequations}
\begin{eqnarray}
& & \max_{\{p_i, w_i\}} \quad \sum_{i\in \mathcal{N}} w_i \log \left( 1 + \frac{h_i p_i}{w_i} \right) \label{pp1.1_1} \\
& & {\rm s.t.} \ \sum_{i\in \mathcal{N}} p_i \leq p \label{pp1.1_2} \\
& & \quad \ \ \sum_{i\in \mathcal{N}} w_i \leq w \label{pp1.1_3}
\end{eqnarray}
\end{subequations}
which is denoted by $\{p^{\ast}_i|i \in \mathcal{N}\}$, is $p^{\ast}_k = p$, $w^{\ast}_k = w$, and $p^{\ast}_i = w^{\ast}_i =
0$, $\forall i \neq k$, where $k = {\rm arg} \, \max_{i \in \mathcal{N}}h_i$. }

\textit{Proof of Lemma~4:} Consider if $\mathcal{N} = \{1,2\}$. Then the problem \eqref{pp1.1_1}--\eqref{pp1.1_3} is equivalent
to
\begin{equation}\label{pp1.2}
\max_{p_1 \leq p, \; w_1 \leq w} \quad g(w,p) = w \log \left( 1 + \frac{h_1p_1}{w_1} \right) +
(w - w_1) \log \left( 1 + \frac{h_2(p - p_1)}{w - w_1} \right) .
\end{equation}
Assume without loss of generality that $h_1 > h_2$. Consider if the constraints $0 \leq p_1 \leq p$ and $0 \leq w_1 \leq w$ are
inactive at optimality. Since the problem \eqref{pp1.2} is convex, using the Karush-Kuhn-Tucker (KKT) conditions, we have
\begin{subequations}
\begin{eqnarray}\label{pp1.3}
& & \log \left( 1 + \frac{h_1p^{\ast}_1}{w^{\ast}_1} \right) -
\frac{h_1p^{\ast}_1} {w^{\ast}_1 + h_1 p^{\ast}_1} - \log \left(1
+ \frac{h_2 (p - p^{\ast}_1)}{w - w^{\ast}_1} \right) +
\frac{h_2(p - p^{\ast}_1)} {w - w^{\ast}_1 + h_2(p - p^{\ast}_1)}
\nonumber \\
& & \qquad \qquad \qquad = y \left(
\frac{h_1p^{\ast}_1}{w^{\ast}_1} \right) - y \left( \frac{h_2(p -
p^{\ast}_1)}{w - w^{\ast}_1} \right) = 0
\end{eqnarray}
\begin{equation}\label{pp1.5}
\frac{h_1w^{\ast}_1} {w^{\ast}_1+h_1p^{\ast}_1} - \frac{h_2(w - w^{\ast}_1)}{w - w^{\ast}_1 + h_2(p - p^{\ast}_1)} = 0.
\end{equation}
\end{subequations}
where $y(x) \triangleq \log(1+x) - x/(1+x)$. Since $y(x)$ is monotonically increasing, it can be seen from \eqref{pp1.3} that
\begin{equation}\label{pp1.4}
\frac{h_1p^{\ast}_1}{w^{\ast}_1} = \frac{h_2(p - p^{\ast}_1)}{w - w^{\ast}_1}.
\end{equation}
Combining \eqref{pp1.5} and \eqref{pp1.4}, we obtain $h_1 = h_2$, which contradicts the condition $h_1 > h_2$. Therefore, at
least one of the constraints $0 \leq p_1 \leq p$ and $0 \leq w_1 \leq w$ is active at optimality. Then it can be shown that
$p^{\ast}_1 = p$ and $w^{\ast}_1 = w$. Note that this is also the optimal solution if $h_1 = h_2$ is assumed. Furthermore, this
conclusion can be directly extended to the case of $N > 2$ by induction. This completes the proof. $\hfill\square$

Now we are ready to show Proposition~1. It can be seen from Lemma~1 that ${P^{\rm S}_i}^{\ast} = {P^{\rm S}_i}^{\star}$,
$\forall i \in \mathcal{I}$, and ${P^{\rm S}_i}^{\ast} = 0$, $\forall i \notin \mathcal{I}$. Then the problem
\eqref{sum1}--\eqref{sum3} is equivalent to
\begin{subequations}
\begin{eqnarray}
& &  \max_{\{W^{\rm S}_i\}} \; \sum_{i\in \mathcal{I}} W^{\rm S}_i
\log \left( 1 + \frac{{P^{\rm S}_i}^{\star} h^{\rm SD}_i}
{W^{\rm S}_iN_0} \right) \label{pp3.3_1} \\
& & {\rm s.t.} \ \sum_{i\in \mathcal{I}} W^{\rm S}_i
\leq W. \label{pp3.3_2}
\end{eqnarray}
\end{subequations}
Since the problem \eqref{pp3.3_1}--\eqref{pp3.3_2} is convex , using the KKT conditions, we have
\begin{subequations}
\begin{equation}\label{pp3.4_1}
\log \left( 1 + \frac{{P^{\rm S}_i}^{\star} h^{\rm SD}_i} {{W^{\rm S}_i}^{\ast}N_0} \right) - \frac{{P^{\rm S}_i}^{\star} h^{\rm SD}_i}
{{W^{\rm S}_i}^{\ast} N_0 + {P^{\rm S}_i}^{\star} h^{\rm SD}_i} - \lambda^{\ast} = y \left(\frac{{P^{\rm S}_i}^ {\star} h^{\rm SR}_i}
{{W^{\rm S}_i}^{\ast}N_0}\right) - \lambda^{\ast} = 0, \; i \in \mathcal{I}
\end{equation}
\begin{equation}\label{pp3.4_2}
W - \sum_{i \in \mathcal{I}} {W^{\rm S}_i}^{\ast} = 0
\end{equation}
\end{subequations}
where $\lambda^{\ast}$ denotes the optimal Lagrange multiplier, and $y(x) \triangleq \log(1+x) - x/(1+x)$. Since $y(x)$ is
monotonically increasing, it follows from \eqref{pp3.4_1} that
\begin{equation}\label{pp3.5}
\frac{{P^{\rm S}_i}^{\star} h^{\rm SR}_i} {{W^{\rm S}_i}^{\ast}N_0} =
\frac{{P^{\rm S}_j}^{\star} h^{\rm SR}_j} {{W^{\rm S}_j}^{\ast}N_0}, \quad
\forall i,j \in \mathcal{I}_1, \ i \neq j.
\end{equation}
Solving the system of equations \eqref{pp3.4_2} and \eqref{pp3.5}, we obtain ${W^{\rm S}_i}^{\ast} = W h^{\rm SD}_i {P^{\rm
S}_i}^{\star} / \sum_{j \in \mathcal{I}} h^{\rm SD}_j{P^{\rm S}_j}^{\star}$, $i\in \mathcal{I}$. This completes the
proof.$\hfill\square$

\textit{Proof of Proposition~2:} \ It can be seen that
\begin{equation}\label{pp2}
C^{\rm SD}_i + C^{\rm SD}_j = \min\{C^{\rm SR}_i, C^{\rm RD}_i\} + \min \{ C^{\rm SR}_j, C^{\rm RD}_j \}
\leq \min \{C^{\rm SR}_i + C^{\rm SR}_j, C^{\rm RD}_i
+ C^{\rm RD}_j\}
\end{equation}
When $P^{\rm S}_j = W^{\rm S}_j = P^{\rm R}_j = W^{\rm R}_j = 0$, it follows from Lemma~1 that the maximum value of the right
hand side of \eqref{pp2} is achieved and equals to $C^{\rm SD}_i$ and, on the other hand, the left hand side of \eqref{pp2}
also equals to $C^{\rm SD}_i$. Therefore, the maximum value of $C^{\rm SD}_i + C^{\rm SD}_j$ is achieved when $P^{\rm S}_j =
W^{\rm S}_j = P^{\rm R}_j = W^{\rm R}_j = 0$. This completes the proof. $\hfill\square$

\section*{Appendix B: Proofs of Lemmas, Propositions, and Theorems in Section~\ref{sec_ad} }

\textit{Proof of Proposition~4:} It is equivalent to show that there exists a feasible point $\{P^{\rm S}_i,W^{\rm S}_i|i \in
\mathcal{I}\}$ of the problem \eqref{ad1}--\eqref{ad3} if and only if $G(\mathcal{I}) \leq W$. If $\{P^{\rm S}_i,W^{\rm S}_i|i
\in \mathcal{I}\}$ is a feasible point of the problem \eqref{ad1}--\eqref{ad3}, then since it is also a feasible point of the
problem \eqref{G1}--\eqref{G3}, we have $G(\mathcal{I}) \leq \sum_{i \in \mathcal{I}} W^{\rm S}_i \leq W$. If we have
$G(\mathcal{I}) \leq W$, then the optimal solution of the problem \eqref{G1}--\eqref{G3} for $\mathcal{I}$, denoted by
$\{{P^{\rm S}_i}^{\ast},{W^{\rm S}_i}^{\ast}|i \in \mathcal{I}\}$, is a feasible point of the problem \eqref{ad1}--\eqref{ad3}
since $\sum_{i \in \mathcal{I}} {W^{\rm S}_i}^{\ast} = G(\mathcal{I}) \leq W$. This completes the proof. $\hfill\square$

\textit{Proof of Theorem~1:} We first show that C1 and C2 are sufficient conditions.

Define $V(n) \triangleq G(\mathcal{N} {(t-1)}) - G(\mathcal{N} {(t)})$ for $n = n {(t)}$, $1 \leq t \leq N$. It follows from C2
that $V(\bar{n}_{S_i} {(1)}) > V(\bar{n}_{S_i} {(2)}) > \cdots > V(\bar{n}_{S_i} {(N_{S_i})})$, $\forall i\in \mathcal{M}$.
Then using Proposition~2, we have $n{(t)}=\textrm{arg}\max_{n \in \mathcal{N} {(t-1)}}V(n)$, $1 \leq t \leq N$. Therefore, we
obtain
\begin{equation}\label{tm1}
V(n {(1)}) > V(n {(2)}) > \cdots > V(n {(N)}).
\end{equation}

It can be seen from C1 that $\mathcal{N} \setminus \mathcal{N}^{\ast}_{N - t} \cap \mathcal{N}_{S_i} = \textrm{arg}
\min_{\mathcal{I} \subseteq \mathcal{N}_{S_i}, |\mathcal{I}| = t_i} G(\mathcal{N}_{S_i} \setminus \mathcal{I}) =
\{\bar{n}_{S_i} {(j)}|1 \leq j \leq t_i\}$, $\forall i\in \mathcal{M}$, where $t_i \triangleq |\mathcal{N} \setminus
\mathcal{N}^{\ast}_{N - t} \cap \mathcal{N}_{S_i}|$. Then we have $\mathcal{N} \setminus \mathcal{N}^{\ast}_{N - t} =
\{\bar{n}_{S_i} {(j)}|1 \leq j \leq t_i, i \in \mathcal{M}\}$ and $G(\mathcal{N}) - G(\mathcal{N}^{\ast}_{N - t}) = \sum_{i \in
\mathcal{M}} \sum^{t_i}_{j=1} V(\bar{n}_{S_i} {(j)})$. Therefore, we obtain $\{t_i|i \in \mathcal{M}\} = \textrm{arg}
\max_{\{k_i\}; \sum_{i \in \mathcal{M}} k_i = t} \sum_{i \in \mathcal{M}} \sum^{k_i}_{j=1} V(\bar{n}_{S_i} {(j)})$. Since it
follows from C2 that $V(\bar{n}_{S_i} {(1)}) > V(\bar{n}_{S_i} {(2)}) > \cdots > V(\bar{n}_{S_i} {(N_{S_i})})$, $\forall i\in
\mathcal{M}$, we have $\mathcal{N} \setminus \mathcal{N}^{\ast}_{N - t} = \textrm{arg} \max_{\mathcal{I} \in \mathcal{N},
|\mathcal{I}| = t} \sum_{n \in \mathcal{I}} V(n) = \{n{(i)}|1 \leq i \leq t\} = \mathcal{N} \setminus \mathcal{N}{(t)}$, where
the second equality is from \eqref{tm1}. This completes the proof for sufficiency of C1 and C2.

We next show that C1 and C2 are necessary conditions by giving two instructive counter examples.

Consider if C1 does not hold. Assume without loss of generality that $\mathcal{M} = \{1\}$. Then it can be seen that C1 is
equivalent to the condition \eqref{opt} and, therefore, the condition \eqref{opt} does not hold, either.

Consider if C2 does not hold. Assume without loss of generality
that $\mathcal{M} = \{2\}$, $N_{S_2} = 1$ and
$G(\bar{\mathcal{N}}_{S_1} {(1)}) - G(\bar{\mathcal{N}}_{S_1}
{(2)}) > G(\mathcal{N}_{S_2}) - G(\bar{\mathcal{N}}_{S_2} {(1)})
> G(\mathcal{N}_{S_1}) - G(\bar{\mathcal{N}}_{S_1} {(1)})$. Then
we have $\mathcal{N}^{\ast}_{N - 2} = \mathcal{N} \setminus
\{\bar{n}_{S_1} {(1)},\bar{n}_{S_1} {(2)}\}$, while it follows
from Proposition~2 that $\mathcal{N}{(2)} = \mathcal{N} \setminus
\{\bar{n}_{S_1} {(1)},\bar{n}_{S_2} {(1)}\}$. Therefore,
$\mathcal{N}^{\ast}_{N - 2} \neq \mathcal{N}{(2)}$. This completes
the proof for necessity of C1 and C2. $\hfill\square$

\textit{Proof of Proposition~6:} The proof of this proposition is built upon the following two lemmas. It suffices to show that
C2 holds for $i = 1$.

\textbf{Lemma 5:} {\it If $p_1 > p_2 > \Delta p > 0$, the following inequality holds}
\begin{equation}\label{lm2.1}
F_i(p_1 - \Delta p) - F_i(p_1) < F_i (p_2 - \Delta p) - F_i (p_1).
\end{equation}

\textit{Proof of Lemma~5:} It can be shown that $F_i(p)$ is a strictly convex and decreasing function of $p$. Using the first
order convexity condition, we have
\begin{equation}\label{lm2.2}
F_i(p_2 - \Delta p) - F_i(p_2) > -F^{\prime}_i(p_2) \Delta p
\end{equation}
and
\begin{equation}\label{lm2.3}
F_i(p_1 - \Delta p) - F_i(p_1) < -F^{\prime}_i(p_1 - \Delta p)
\Delta p
\end{equation}
where $F^{\prime}_i$ is the first order derivative of $F_i$.
Consider two cases. (i)~If $p_2 \leq p_1 - \Delta p$, then
$F^{\prime}_i (p_2) \leq F^{\prime}_i (p_1 - \Delta p)$ due to the
convexity of $F_i (p_2)$. Therefore, using $\Delta p > 0$ together
with \eqref{lm2.2} and \eqref{lm2.3}, we obtain \eqref{lm2.1};
(ii)~If $p_2 \geq p_1 - \Delta p$, using $p_1 > p_2$ and a similar
argument as in 1), we can show that $F_i(p_2) - F_i(p_1) < F_i
(p_2 - \Delta p) - F_i (p_1 - \Delta p)$, which is equivalent to
\eqref{lm2.1}. This completes the proof. $\hfill\square$

$G(\mathcal{N}_{S_1})$ can be extended to $G(\mathcal{N}_{S_1},
P_{S_1})$ if $P_{S_1}$ is considered as a variable.

\textbf{Lemma 6:} {\it $p^{\ast}_i$, $\forall i \in \mathcal{N}_{S_1}$, is increasing with $P_{S_1}$, where $\{p^{\ast}_i|i \in
\mathcal{N}_{S_1}\}$ denotes the optimal solution of the problem \eqref{Gs1}--\eqref{Gs2} for $\mathcal{N}_{S_1}$ and
$P_{S_1}$.}

\textit{Proof of Lemma~6:} The inverse function of $w = F_i (p)$ is $p = F^{-1}_i(w) = (e^{c_i/w} - 1) w / h_i$. Then we have
\begin{subequations}
\begin{eqnarray}
& & G(\mathcal{N}_{S_1}, P_{S_1}) = \max_{w_i} \sum_{i\in \mathcal{N}_{S_1}} w_i \\ \label{lm3.1_1}
& & \textrm{s.t.} \ \sum_{i\in \mathcal{N}_{S_1}} F^{-1}_i(w_i) \leq
P_{S_1}. \label{lm3.1_2}
\end{eqnarray}
\end{subequations}
Since the problem \eqref{lm3.1_1}--\eqref{lm3.1_2} is convex, using the KKT conditions, the optimal solution and the optimal
Lagrange multiplier of this problem, denoted by $\{w^{\ast}_i|i \in \mathcal{N}_{S_1}\}$ and $\lambda^{\ast}$, respectively,
satisfy the following equations
\begin{equation}\label{lm3.2}
1 + \frac{\lambda^{\ast}} {h_i} \left( e^{\frac{c_i}
{w_i^{\ast}}} \left( \frac{c_i}{w_i^{\ast}} - 1 \right) + 1
\right) = 0, \quad \forall i\in \mathcal{N}_{S_1}.
\end{equation}
It can be shown that $(e^{c_i/w_i^{\ast}} (c_i/w_i^{\ast} - 1) + 1) / h_i$ is monotonically decreasing with $w_i^{\ast}$.
Therefore, $w^{\ast}_i$, $\forall i\in \mathcal{N}_{S_1}$, and, correspondingly, $p^{\ast}_i = F^{-1}_i (w^{\ast}_i)$, $\forall
i\in \mathcal{N}_{S_1}$, is decreasing and increasing, respectively, with $\lambda^{\ast}$. Then it follows from
\eqref{lm3.1_2} that $p^{\ast}_i$, $\forall i \in \mathcal{N}_{S_1}$, is increasing with $P_{S_1}$. This completes the proof.
$\hfill\square$

We are now ready to prove this proposition. Let $P_1 > P_2$ and $\mathcal{N}^{-k}_{S_1} \triangleq \mathcal{N}_{S_1} \setminus
\{k\}$ for some $k \in \mathcal{N}_{S_1}$. Let $\{p^{\star}_i|i \in \mathcal{N}^{-k}_{S_1}\}$ denote the optimal solution of
the problem \eqref{Gs1}--\eqref{Gs2} for $\mathcal{N}^{-k}_{S_1}$ and $P_2$. Using Lemma~6, the optimal solution of the problem
\eqref{Gs1}--\eqref{Gs2} for $G(\mathcal{N}_{S_1}, P_2)$ can be expressed as $\{p^{\star}_i - \Delta p_i\}$, $i \in
\mathcal{N}^{-k}_{S_1}$, and $p^{\star}_k$, respectively, where $\Delta p_i > 0$ and $\sum_{i \in \mathcal{N}^{-k}_{S_1}}
\Delta p_i = p^{\star}_k$. Then we have
\begin{equation}\label{pp6.1}
G(\mathcal{N}_{S_1}, P_2) - G(\mathcal{N}^{-k}_{S_1},
P_2) = \sum_{i \in \mathcal{N}^{-k}_{S_1}} \left( F_i
(p^{\star}_i - \Delta p_i) - F_i (p^{\star}_i) \right) + F_k
(p^{\star}_k).
\end{equation}
Let $\{p^{+}_i|i \in \mathcal{N}^{-k}_{S_1}\}$ denote the optimal solution of the problem \eqref{Gs1}--\eqref{Gs2} for
$\mathcal{N}^{-k}_{S_1}$ and $P_1$. Then we have
\begin{eqnarray}\label{pp6.2}
G(\mathcal{N}_{S_1}, P_1) - G(\mathcal{N}^{-k}_{S_1}, P_1)
\!\!\!\!&=&\!\!\!\! \min_{\{p_i\}; \sum_{i\in \mathcal{N}_{S_1}}
p_i \leq P_1} \sum_{i \in \mathcal{N}_{S_1}} F_i (p_i) - \sum_{i
\in \mathcal{N}^{-k}_{S_1}} F_i (p^{+}_i) \nonumber \\
\!\!\!\!&\leq &\!\!\!\! \sum_{i \in \mathcal{N}^{-k}_{S_1}}(F_i
(p^{+}_i - \Delta p_i ) - F_i(p^{+}_i)) + F_k(p^{\star}_k).
\end{eqnarray}
Since $P_1 > P_2$, it follows from Lemma~6 that $p^{+}_i > p^{\star}_i > \Delta p_i > 0$, $i \in \mathcal{N}^{-k}_{S_1}$. Using
Lemma~5, we obtain $F_i(p^{+}_i - \Delta p_i) - F_i(p^{+}_i) < F_i(p^{\star}_i - \Delta p_i) - F_i(p^{\star}_i)$, $j\in
\mathcal{N}^{-k}_{S_1}$. Therefore, comparing \eqref{pp6.1} with \eqref{pp6.2}, we have
\begin{equation}\label{pp6.3}
G(\mathcal{N}_{S_1}, P_1) - G(\mathcal{N}^{-k}_{S_1}, P_1) < G(\mathcal{N}_{S_1}, P_2) - G(\mathcal{N}^{-k}_{S_1}, P_2).
\end{equation}
which can be rewritten as
\begin{equation}\label{pp6.4}
G(\mathcal{N}^{-k}_{S_1}, P_2) - G(\mathcal{N}^{-k}_{S_1}, P_1) < G(\mathcal{N}_{S_1}, P_2) - G(\mathcal{N}_{S_1}, P_1).
\end{equation}

Let $\{p^{\ast}_i|i \in \mathcal{N}_{S_1}\}$, denote the optimal solution of the problem \eqref{Gs1}--\eqref{Gs2} for
$\mathcal{N}_{S_1}$ and $P_{S_1}$. Then we have
\begin{eqnarray}\label{pp6.5}
G(\mathcal{N}_{S_1} \!\setminus\! \{\bar{n}_{S_1}{(2)}\}, P_{S_1})
\!-\! G(\bar{\mathcal{N}}_{S_1}{(2)}, P_{S_1}) \!\!\!\!& \leq
&\!\!\!\! F_{\bar{n}_{S_1}{(1)}} (p^{\ast}_{\bar{n}_{S_1}{(1)}})
\!+\! G(\bar{\mathcal{N}}_{S_1} {(2)}, P_{S_1} \!-\!
p^{\ast}_{\bar{n}_{S_1}{(1)}}) \!-\!
G(\bar{\mathcal{N}}_{S_1}{(2)}, P_{S_1}) \nonumber \\
\!\!\!\! &<&\!\!\!\!  F_{\bar{n}_{S_1}{(1)}}
(p^{\ast}_{\bar{n}_{S_1}{(1)}}) \!+\!
G(\bar{\mathcal{N}}_{S_1}{(1)},
P_{S_1} \!-\! p^{\ast}_{\bar{n}_{S_1}{(1)}}) \!-\!
G(\bar{\mathcal{N}}_{S_1}{(1)}, P_{S_1}) \nonumber \\
\!\!\!\!& =&\!\!\!\! G(\mathcal{N}_{S_1}, P_{S_1}) -
G(\bar{\mathcal{N}}_{S_1} {(1)}, P_{S_1})
\end{eqnarray}
where the second inequality follows from \eqref{pp6.4}. On the other hand, we have
\begin{equation}\label{pp6.6}
\begin{split}
G(\mathcal{N}_{S_1} \setminus \{\bar{n}_{S_1}{(2)}\}, P_{S_1}) - G(\bar{\mathcal{N}}_{S_1}{(2)}, P_{S_1})
& \geq G(\mathcal{N}_{S_1} \setminus \{\bar{n}_{S_1}{(1)}\}, P_{S_1}) - G(\bar{\mathcal{N}}_{S_1}{(2)}, P_{S_1})\\
& = G(\bar{\mathcal{N}}_{S_1}{(1)}, P_{S_1}) - G(\bar{\mathcal{N}}_{S_1}{(2)}, P_{S_1}).
\end{split}
\end{equation}
Therefore, comparing \eqref{pp6.5} with \eqref{pp6.6}, we completes the proof. $\hfill\square$

\textit{Proof of Lemma~2:} Assume $\mathcal{N}_{S_l,k} \neq \mathcal{N}^{\ast}_{S_l,k}$. Then there exist $a \in
\mathcal{N}^{\ast}_{S_l,k}$ and $b \in \mathcal{N} \setminus \mathcal{N}^{\ast}_{S_l,k}$ such that $F_a(p) > F_b(p)$. Let
$\{p^{\ast}_i|i \in \mathcal{N}^{\ast}_{S_1,k}\}$ denote the optimal solution of the problem \eqref{Gs1}--\eqref{Gs2} for
$G(\mathcal{N}^{\ast}_{S_l,k})$. Then there always exists $\mathcal{N}^{\prime}_{S_l,k} \triangleq \mathcal{N}^{\ast}_{S_l,k}
\cup \{b\} \setminus \{a\}$ such that
\begin{equation}
\begin{split}
G(\mathcal{N}^{\ast}_{S_l,k})& = \sum_{i \in \mathcal{N}^{\ast}_{S_l,k}, \; i \neq a} F_i(p^{\ast}_i) + F_a (p^{\ast}_a)
> \sum_{i \in \mathcal{N}^{\ast}_{S_l,k}, \; i \neq a} F_i(p^{\ast}_i) + F_b(p^{\ast}_a) \\
& \geq \min_{\{p_i\}; \; \sum_{i \in \mathcal{N}^{\prime}_{S_l,k}} p_i \leq P_{S_1}} \sum_{i \in \mathcal{N}^{\prime}_{S_l,k}}
F_i(p_i) = G(\mathcal{N}^{\prime}_{S_l,k})
\end{split}
\end{equation}
which contradicts the definition of $\mathcal{N}^{\ast}_{S_l,k}$. Then it follows that $\mathcal{N}_{S_l,k} =
\mathcal{N}^{\ast}_{S_l,k}$. Using similar arguments, it can be shown that $\mathcal{N}_{S_l,k} = \bar{\mathcal{N}}_{N_{S_l}}
{(N_{S_l} - k)}$. This completes the proof. $\hfill\square$

\textit{Proof of Proposition~7:} It suffices to show that C1 holds for $i = 1$ if for any $j \in \mathcal{N}_{S_1}$, there
exists no more than one $k \in \mathcal{N}_{S_1}$, $k \neq j$, such that C3 holds. It can be seen that for any $1 \leq k \leq
N_{S_1}$, only two cases are under consideration: (i) there exists $\mathcal{N}_{S_1,k}$ that satisfies the condition given in
Lemma~4 and, therefore, $\mathcal{N}^{\ast}_{S_1,k} = \mathcal{N}{(N_{S_1} - k)}$; (ii) there exist $\mathcal{N}_{S_1,k - 1}$
and $\mathcal{N}_{S_1,k + 1}$ that satisfy the condition given in Lemma~2 respectively and, therefore,
$\mathcal{N}^{\ast}_{S_1, k - 1} = \mathcal{N}{(N_{S_1} - k + 1)} \subseteq \mathcal{N}{(N_{S_1} - k - 1)} =
\mathcal{N}^{\ast}_{S_1, k + 1}$. Then it follows that $\mathcal{N}^{\ast}_{S_1,k} = \mathcal{N}{(N_{S_1} - k)}$. This
completes the proof. $\hfill\square$

\textit{Proof of Lemma~3:} Consider if $F_i(p)$ intersects $F_j(p)$ at a point $(p^{\prime}, w^{\prime})$. Then we obtain
\begin{equation}\label{pp4.1}
\frac{c_j}{c_i} = \frac{w^{\prime} \log \left( 1 + \frac{h_j
p^{\prime}} {w^{\prime}} \right)}{w^{\prime} \log \left( 1 +
\frac{h_i p^{\prime}} {w^{\prime}} \right)} = q \left(
\frac{p^{\prime}}{w^{\prime}} \right)
\end{equation}
where $q(x) = \log (1 + h_j x)/\log (1 + h_i x)$, $0 < x <
\infty$. It can be shown that $\underset{x \rightarrow 0}{\lim}
q(x) = h_j/h_i$, $\underset{x \rightarrow \infty}{\lim} q(x) = 1$,
and $q(x)$ is monotonically decreasing with $x$. Therefore, the
range of $q(x)$ is $(1, h_j/h_i)$. If $c_j/c_i \in (1, h_j/h_i)$,
there exists a unique solution $x^{\prime}$ such that
$q(x^{\prime}) = c_j/c_i$. Hence, $F_i(p)$ and $F_j(p)$ have a
unique intersection point given by $w^{\prime} = c_j/\log (1 + h_j
x^{\prime})$, $p^{\prime}=w^{\prime}x^{\prime}$, and the claim~(i)
follows. If $c_j/c_i \notin (1, h_j/h_i)$, there is a special case
that $F_i(p) = F_j(p)$, $\forall p > 0$ if $h_j/h_i = c_j/c_i =
1$. Otherwise, the solution of \eqref{pp4.1} does not exist, i.e.,
$F_i(p)$ does not intersect $F_j(p)$ and, therefore, the
claims~(ii) and~(iii) also follow. This completes the proof.
$\hfill\square$

\newpage

\end{document}